\documentclass[aps,pre,twocolumn,groupedaddress]{revtex4-2}
\newcommand{\bec}[1]{\mbox{\boldmath $ #1$}}
\usepackage{bm}
\usepackage{graphicx}
\usepackage{color}

\newcommand{\meanUU}{\overline{\bm U}}
\newcommand{\meanU}{\overline{U}}
\newcommand{\meanWW}{\overline{\bm W}}

\newcommand{\meanT}{\overline{T}}
\newcommand{\meanP}{\overline{P}}
\newcommand{\meanTheta}{\overline{\Theta}}
\newcommand{\meanrho}{\overline{\rho}}

\begin{document}
\title{Large-scale circulations in a shear-free convective turbulence: Mean-field simulations}
\author{G. Orian$^{1,2}$}
\author{A. Asulin$^1$}
\author{E. Tkachenko$^1$}
\author{N. Kleeorin$^1$}
\author{A. Levy$^1$}
\author{I. Rogachevskii$^1$}
\email{gary@bgu.ac.il}
\affiliation{\\
$^1$The Pearlstone Center for Aeronautical Engineering
Studies, Department of Mechanical Engineering,
Ben-Gurion University of the Negev, P.O.Box 653,
Beer-Sheva 84105,  Israel\\
$^2$Nuclear Research Center, Negev POB 9001,
Beer-Sheva 84190, Israel}
\date{\today}
\begin{abstract}
It has been previously shown (Phys. Rev. E {\bf 66}, 066305, 2002)
that a non-rotating turbulent convection with nonuniform large-scale flows contributes to the
turbulent heat flux. As a result, the turbulent heat flux depends explicitly not only
on the gradients of the large-scale temperature,
but it also depends on the gradients of the large-scale velocity.
This is because the nonuniform large-scale flows produce anisotropic
velocity fluctuations which modify the turbulent heat flux.
This effect causes an excitation of a convective-wind instability
and formation of large-scale semi-organised coherent structures (large-scale convective cells).
In the present study, we perform mean-field numerical simulations
of shear-free convection which take into account the modification of the turbulent heat flux
by nonuniform large-scale flows.
We use periodic boundary conditions in horizontal direction,
as well as stress-free or no-slip boundary conditions in vertical direction.
We show that the redistribution of the turbulent heat flux by the
nonuniform large-scale motions in turbulent convection
plays a crucial role in the formation of the large-scale semi-organised coherent structures.
In particular, this effect results in a strong reduction of the critical effective Rayleigh number
(based on the eddy viscosity and turbulent temperature diffusivity)
required for the formation of the large-scale convective cells.
We demonstrate that the convective-wind instability is excited when the scale separation ratio
between the height of the convective layer and the integral turbulence scale is large.
The  level of the mean kinetic energy at saturation increases
with the scale separation ratio.
We also show that inside the large-scale convective cells, there
are local regions with the positive vertical gradient of the
potential temperature which implies that these regions are stably stratified.
\end{abstract}

\maketitle


\section{Introduction}
\label{sec-I}

Turbulent non-rotating convection has been investigated in theoretical studies \cite{Z91,C14,S94,AGL09,LX10,T73,MY75,RI21},
laboratory  experiments  \cite{AGL09,LX10,BPA00,ZX10,KH81,SWL89,CCL96,NSS01,NS03,BKT03,XSZ03,XL04,SQ04,FA04,BNA05,LLT05,FBA08,EEKR06,BEKR09,BEKR11},
numerical simulations \cite{HTB03,PHP04,R06,S08,CS12,AB16,KRB17,KA19,SS20,PSS21}, and atmospheric observations \cite{KF84,EB93,AZ96,ZGH98,YKH02}.
One of the key features of turbulent convection
is the formation of coherent semi-organised structures
referred as large-scale convective cells (or large-scale circulations)
in a shear-free convection and large-scale rolls in a sheared convection.
Spatial scales of the large-scale coherent structures in a turbulent
convection are much larger than turbulent scales and their life-time
is much longer than the characteristic turbulent timescale.

Buoyancy-driven structures, such as plumes, jets, and large-scale
circulation patterns are observed in numerous laboratory
experiments on turbulent convection.
The large-scale circulations caused by convection in the
Rayleigh-B{\'e}nard apparatus are often called the ''mean wind" \citep{KH81}.
There are several open questions concerning these flows, e.g., how
do they arise, and what are their characteristics and dynamics.

In atmospheric turbulent convection two types of coherent semi-organised
structures, namely ''cloud cells" in a shear-free convection
and ''cloud streets" in a sheared convection have been observed \cite{EB93,AZ96,YKH02}.
In particular, cloud cells are seen as
three-dimensional, long-lived B\'{e}nard-type cells
consisting in narrow uprising plumes and wide downdraughts. These structures
are occupied the entire convective boundary layer of about 1-3 km in height.
In the sheared convective boundary layer with a strong wind, the cloud streets
are seen as large-scale rolls stretched along the wind.

In spite of a number of theoretical and numerical studies,
the nature of large-scale coherent
structures in turbulent convection is a subject of active discussions.
There are two points of view on the
origin of large-scale circulation in turbulent convection.
According to one point of view, the large-scale circulations which develop at low
Rayleigh numbers near the onset of convection
continually increase their size as the Rayleigh numbers is increased and
continue to exist in an average sense at even very high Rayleigh
numbers \cite{HTB03}.
Another hypothesis holds that the large-scale circulation is a
genuine high Rayleigh number effect \citep{KH81}.

A mean-field theory of non-rotating turbulent
convection has been developed in Refs. \cite{EKRZ02,EKRZ06},
where the convective-wind instability in the shear-free turbulent
convection causing the formation of large-scale cells
is identified.
In the sheared turbulent convection, the convective-shear instability
resulting in formation of large-scale rolls can be excited
\cite{EKRZ02,EKRZ06}.
A redistribution of the turbulent heat flux by non-uniform large-scale motions plays a
crucial role in the formation of the large-scale coherent structures
in turbulent convection.
To determine key parameters that affect formation of the large-scale
coherent structures in the turbulent convection,
the linear stage of the convective-wind and convective-shear instabilities have been
numerically studied in Ref. \cite{EGKR06}.

In the present study we perform mean-field numerical simulations to study nonlinear evolution
of large-scale circulations in turbulent shear-free convection
taking into account the effect of modification of the
turbulent heat flux by non-uniform large-scale motions.
This paper is organized as follows.
In Section~II we discuss a modification of the turbulent heat flux caused
by anisotropic velocity fluctuations in turbulence with non-uniform large-scale flows.
In Section~III we formulate the non-dimensional equations, the governing non-dimensional parameters
and discuss the large-scale convective-wind instability.
In Section~IV we describe the set-up for the mean-field numerical simulations,
and discuss the results of the numerical simulations for periodic
boundary conditions in horizontal direction,
as well as stress-free or no-slip boundary conditions in vertical direction.
In Section~V we discuss the novelty aspects and significance of the obtained results,
as well as their applications.
Finally, conclusions are drawn in Section~VI.

\section{Turbulent convection and turbulent heat flux}
\label{sec-II}

We consider turbulent convection with very high Rayleigh numbers, and large
Reynolds and Peclet numbers.
Formation of coherent semi-organised structures is usually studied
using a mean field approach whereby the velocity ${\bm U}$, pressure $P$ and
potential temperature $\Theta$ are decomposed into the mean and fluctuating parts:
${\bm U} = \meanUU + {\bm u}$, $P = \meanP + p$ and $\Theta =
\meanTheta + \theta$, the fluctuating parts have zero mean values,
which implies the Reynolds averaging.
Here $\meanUU = \langle {\bm U} \rangle$ is the mean velocity, $\meanP = \langle P
\rangle$ is the mean pressure and  $\meanTheta = \langle \Theta \rangle$
is the mean potential temperature, and ${\bm u}$, $p$ and $\theta$
are fluctuations of velocity, pressure and potential temperature, respectively.
Averaging the Navier-Stokes equation and equation for the potential temperature
over an ensemble, we obtain the mean-field equations:
\begin{eqnarray}
\biggl({\partial  \over \partial t} + \meanUU \cdot
\bec{\nabla}\biggr) \meanU_{i} &=& - {\nabla}_{i} \biggl({\meanP
\over \meanrho_0}\biggr) - {\nabla}_{j} \langle u_{i} \, u_{j} \rangle + \beta \, \meanTheta e_i,
\nonumber \\
\label{LB1}
\end{eqnarray}
\begin{eqnarray}
\biggl({\partial  \over \partial t} + \meanUU \cdot
\bec{\nabla} \biggr) \meanTheta &=& - (\meanUU \cdot
\bec{\nabla}) \meanT_0 -{\nabla}_{i} \langle u_{i} \theta \rangle ,
\label{LB2}
\end{eqnarray}
and ${\rm div} \, \meanUU= 0$, where the mean potential temperature $\meanTheta$ is related to the physical temperature $\meanT$ as: $\meanTheta=\meanT \,(\meanP_0/\meanP)^{1-1/\gamma}$, where $\meanT$ is the mean physical temperature
and $\meanT_0$ is the mean physical temperature in the equilibrium (the basic reference state),
$\meanP$ is the mean pressure and $\meanP_0$  is the mean pressure in the equilibrium, $\meanrho_0$ is the mean fluid density in the equilibrium, $\gamma=c_p/c_v$ is the specific heats ratio,
$\beta=g/\meanT_0$ is the buoyancy parameter, ${\bf g}$  is the gravity acceleration, and $\meanrho_0$ is the mean fluid density.
Equations~(\ref{LB1})--(\ref{LB2}) are written in the Boussinesq approximation with ${\rm div} \, \meanUU= 0$.

In Eqs.~(\ref{LB1})--(\ref{LB2}) we neglect very small terms caused by kinematic viscosity and molecular diffusivity of temperature in comparison with the terms due to turbulent viscosity and turbulent diffusivity.
The mean velocity $\meanUU(t, {\bm x})$, the mean potential temperature
$\meanTheta(t, {\bm x})$ and the mean pressure $\meanP(t, {\bm x})$ in Eqs.~(\ref{LB1})--(\ref{LB2}) describe deviations from
the hydrostatic equilibrium without mean motions: ${\nabla} \meanP_{0} = \meanrho_0 {\bm g}$ and $\meanrho_0 =$ const, where ${\bm g}=-g \, {\bm e}$ and ${\bm e}$ is the vertical unit vector.

The effect of convective turbulence on the mean velocity and mean potential temperature is described by the Reynolds stress $\langle u_{i} \, u_{j} \rangle$ and turbulent flux of potential temperature
${\bm F} = \langle {\bm u} \theta \rangle$.
Traditional theoretical turbulence models,
such as the Kolmogorov-type local closures, imply that the turbulent flux
of momentum determined by $\langle u_{i} \, u_{j} \rangle$ and
the turbulent flux of potential temperature
${\bm F}$ are assumed to be proportional
to the local mean gradients, whereas the proportionality
coefficients, namely turbulent viscosity and turbulent temperature diffusivity, are
uniquely determined by local turbulent parameters. The classical expression for the Reynolds stress is
$\langle u_{i} \, u_{j} \rangle = - \nu_{_{T}} (\nabla_i \meanU_j + \nabla_j \meanU_i)$
and the classical turbulent heat flux is given by
${\bm F} = - \kappa_{_{T}} \bec{\nabla} \meanTheta$ (see, e.g., Ref.~\cite{MY75}), where
$\nu_{_{T}}$ is the turbulent viscosity and
$\kappa_{_{T}}$ is the turbulent temperature diffusivity.

On the other hand, there are coherent structures in turbulent convection
(large-scale coherent convective cells or large-scale circulations)
and the velocity field inside large-scale circulations
is strongly nonuniform. These nonuniform motions can produce
anisotropic velocity fluctuations which can contribute
to the turbulent heat flux.
In particular, the turbulent heat flux, ${\bm F} = - \kappa_{_{T}} \bec{\nabla} \meanTheta$, does not
take into account the contribution from anisotropic velocity fluctuations.

It has been shown in Ref.~\cite{EKRZ02} that the contribution to the turbulent heat flux
from anisotropic velocity fluctuations plays essential role in
formation of large-scale circulations in turbulent convection.
In particular, the following expression for the
turbulent heat flux ${\bm F}$ has been derived in Ref.~\cite{EKRZ02}:
\begin{eqnarray}
{\bm F} = {\bm F}^{\ast} -  \tau_0 \, \left[{\bm
F}_{z}^{\ast} \, {\rm div} \, \meanUU_{\perp} - {1 \over 2} \,
\meanWW {\bm \times} {\bm F}_z^{\ast}\right] ,
 \label{R100}
\end{eqnarray}
where ${\bm F}^{\ast} = - \kappa_{_{T}} \bec{\nabla} \meanTheta$
is the classical turbulent heat flux (i.e., the background turbulent heat flux
in the absence of nonuniform large-scale flows),
$\tau_0$ is the correlation time of turbulent velocity
at the integral scale of turbulent motions, $\meanWW =
\bec{\nabla} {\bm \times}  \meanUU$ is the mean vorticity, the mean velocity
$\meanUU=  \meanUU_{\perp} +  \meanUU_{z}$ is decomposed into the horizontal
$\meanUU_{\perp}$ and vertical $\meanUU_{z}$
components.
The new terms in the turbulent heat flux ${\bm F}$ are caused by
anisotropic velocity fluctuations and depend on
the mean velocity gradients. These new terms lead
to the excitation of large-scale convective-wind instability and formation of
coherent structures. For nearly uniform large-scale flows, the anisotropic
turbulence effects can be neglected, so that the traditional equation
for the turbulent heat flux is recovered.

\begin{figure}
\vspace*{1mm}
\centering
\includegraphics[width=8cm]{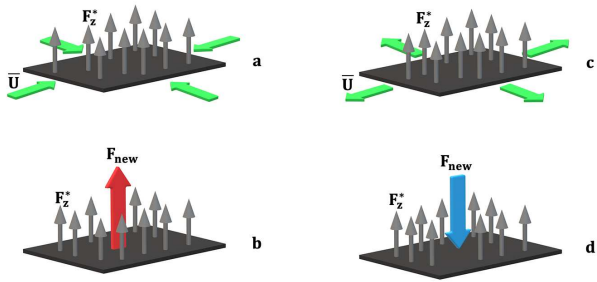}
\caption{\label{Fig1} The illustration of the physics caused by the new turbulent heat flux
${\bm F}_{\rm new} = - \tau_0 \, {\bm F}_{z}^{\ast} \, {\rm div}
\, \meanUU_{\perp}$ produced by the perturbations of the
convergent (or divergent) horizontal mean flows $\meanUU_{\perp}$ (shown by the green arrows
in panels {\bf a} and {\bf c}).
The new turbulent flux ${\bm F}_{\rm new}$ increases the upward (positive) heat flux, enhances buoyancy and increases the local mean potential temperature, thus creating the upward flow. Likewise, the new turbulent flux ${\bm F}_{\rm new}$ decreases the vertical turbulent flux of potential temperature by the divergent horizontal motions, which reduces the buoyancy and decreases the local mean potential temperature, thus creating the downward flow. These effects cause a formation of the large-scale coherent structures (the large-scale circulations).
}
\end{figure}

The physical meaning of the new terms in the turbulent heat flux is the following. The first
term ${\bm F}_{\rm new} = - \tau_0 \, {\bm F}_{z}^{\ast} \, {\rm div}
\, \meanUU_{\perp}$ in the squared brackets of Eq.~(\ref{R100}) for the turbulent heat flux
describes the redistribution of the vertical background turbulent
heat flux ${\bm F}_{z}^{\ast}$ by the perturbations of the
convergent (or divergent) horizontal mean flows $\meanUU_{\perp}$ (see Fig.~\ref{Fig1}).
This redistribution of the vertical turbulent heat flux
occurs during the life-time of turbulent eddies.
In particular, the term ${\bm F}_{\rm new}$  describes enhancing of the vertical turbulent flux of potential temperature by the converging horizontal motions. The latter increases the upward (positive) heat flux and enhances buoyancy, thus creating the upward flow. In its turn, this flow strengthens the horizontal convergent flow causing a large-scale convective-wind instability.

The second term $\propto (\tau_0/2) \, (\meanWW {\bm \times}
{\bm F}_z^{\ast})$ in the squared brackets of Eq.~(\ref{R100}) determines
the formation of the horizontal turbulent heat flux due to
"rotation" of the vertical background turbulent heat flux ${\bm
F}_{z}^{\ast}$  by the perturbations of the horizontal mean
vorticity $\meanWW_{\perp}$.
In particular, this term  describes generation of horizontal turbulent flux of potential temperature through turning the vertical flux by horizontal component of the vorticity. This effect decreases (increases) local potential temperatures in rising motions, that reduces the buoyancy accelerations, and weakens vertical velocity and vorticity, and thus causes damping of large-scale convective-wind instability.
These two effects are determined by the local inertial forces in nonuniform mean flows.

The linear-stability analysis of the linearised Eqs.~(\ref{LB1})--(\ref{R100}) yields the following estimate for the growth rate $\tilde\gamma_{\rm inst}$  of long-wave
perturbations $[(\ell_0 K)^2 \ll 1]$ of the convective-wind instability:
\begin{eqnarray}
\tilde\gamma_{\rm inst} = \left({\beta \, F_{z}^{\ast} \, \tau_0 \over 2}\right)^{1/2} \, K \, |\sin \varphi| \, (1 - 2 \sin^2 \varphi)^{1/2} - \nu_{_{T}} K^2,
\nonumber\\
\label{R200}
\end{eqnarray}
where $\varphi$ is the angle between the vertical axis $z$ and the wave vector ${\bm K}$ of small perturbations and $u_0$ is the characteristic turbulent velocity in the integral scale $\ell_0$ of turbulence.
Equation~(\ref{R200}) is obtained for the turbulent Prandtl number
${\rm Pr}_{_{T}}=1$.
The analysis of the convective-wind instability is performed in Refs.~\cite{EKRZ02,EKRZ06,EGKR06}.
The mechanism of this instability is as follows (see Fig.~\ref{Fig1}).
When $\nabla_z\meanU_{z}>0$, perturbations of the vertical velocity $\meanU_{z}$  cause negative divergence of the horizontal velocity, div$\, \meanU_{\perp}<0$  (provided that  div$\, \meanUU=0$),
describing convergent horizontal flows (shown by the green arrows
in panel {\bf a} of Fig.~\ref{Fig1}). This produces vertical turbulent flux of potential temperature ${\bm F}_{\rm new} = - \tau_0 \, {\bm F}_{z}^{\ast} \, {\rm div} \, \meanUU_{\perp}$ (shown by the red arrow in panel {\bf b} of Fig.~\ref{Fig1}). The latter strengthens the local total vertical turbulent flux of potential temperature and by this means leads to increasing the local mean potential temperature and buoyancy. The latter enhances the local mean vertical velocity  $\meanU_{z}$. Through this mechanism a large-scale convective-wind instability is excited.

Similar reasoning is valid when  $\nabla_z\meanU_{z}<0$, whereas  div$\, \meanU_{\perp}>0$ describing divergent horizontal flows (shown by the green arrows
in panel {\bf c} of Fig.~\ref{Fig1}). This produces negative perturbations of the vertical flux of potential temperature ${\bm F}_{\rm new} = - \tau_0 \, {\bm F}_{z}^{\ast} \, {\rm div} \, \meanUU_{\perp}$ (shown by the blue arrow in panel {\bf d} of Fig.~\ref{Fig1}), which
lead to decrease of the mean potential temperature and buoyancy. This enhances the downward flow, and results in excitation of the convective-wind instability. Thus, nonzero perturbations of div$\, \meanU_{\perp}$   cause redistribution of the vertical turbulent flux of potential temperature and formation of regions with large values of this flux. The regions where $\nabla_z\meanU_{z}<0$  alternate with the low-flux regions where  $\nabla_z\meanU_{z}>0$. This mechanism causes formation of the large-scale circulations.

\section{Nondimensional equations and large-scale convective-wind instability}
\label{sec-III}

Using the expression~(\ref{R100}) for the turbulent heat flux ${\bm F}$ with the additional terms
caused by the nonuniform mean flows,
calculating div ${\bm F}$, and assuming that the nondimensional total vertical heat flux
$\Phi_c=\tilde F_z^\ast + \tilde U_z \, \tilde\Theta$ is constant,
we rewrite Eqs.~(\ref{LB1})--(\ref{LB2}) in a nondimensional form:
\begin{eqnarray}
&& {\partial \tilde{\bm U} \over \partial t} + (\tilde{\bm U} \cdot {\bm \nabla}) \tilde{\bm U} = - {{\bm \nabla} \tilde P \over \rho_0} + {\rm Ra}_{_{T}} \, \tilde\Theta \, {\bm e} + \Delta \tilde{\bm U},
 \label{B1}
\end{eqnarray}
\begin{eqnarray}
&&  {\rm Pr}_{_{T}} \, \left({\partial \tilde\Theta \over \partial t} + (\tilde{\bm U} \cdot {\bm \nabla}) \tilde\Theta\right) =  \tilde U_z +  \Delta \tilde\Theta
+ \epsilon \biggl[ \left(\nabla_z \tilde U_z\right)
\nonumber\\
&& \quad
\times \nabla_z \left( \tilde U_z \, \tilde\Theta \right) + \left(\Phi_c - \tilde U_z \, \tilde\Theta \right) \, \left({\Delta \over 2}-\nabla_z^2\right) \tilde U_z
\nonumber\\
&& \quad
+ {1 \over 2} \, \left(\nabla_z \tilde U_x - \nabla_x \tilde U_z\right) \, \nabla_x \left(\tilde U_z \, \tilde\Theta \right)
\nonumber\\
&& \quad
+ {1 \over 2} \,\left(\nabla_z \tilde U_y - \nabla_y \tilde U_z\right)
 \nabla_y \left(\tilde U_z \, \tilde\Theta \right)
\biggr],
 \label{B2}
\end{eqnarray}
and ${\rm div} \, \tilde{\bm U}= 0$ (see Appendix~\ref{Appendix:A}),
where  $\tilde{\bm U}$ is the nondimensional mean velocity,
$\tilde F_z^\ast$ is the nondimensional vertical turbulent background heat flux,
 $\tilde\Theta$ is the nondimensional mean potential temperature
and  $\tilde P$ is the nondimensional  mean pressure,
and  ${\bm e}$ is the unit vector directed along the vertical $z$ axis.

In Eqs.~(\ref{B1})--(\ref{B2}),  length is measured in the units of the vertical size of the convective layer $L_z$ (e.g., the size of the computational domain), time is measured in the units of the turbulent viscosity time, $L_z^2/\nu_{_{T}}$, velocity is measured in the units of $\nu_{_{T}}/L_z$, potential temperature is measured in the units of $L_z \, N^2 \, {\rm Pr}_{_{T}}/\beta$, turbulent heat flux is measured in the units of $\nu_{_{T}} \, N^2 \, {\rm Pr}_{_{T}}/\beta$ and pressure is measured in the units of $\rho_0 \, (\nu_{_{T}}/L_z)^2$. Here $\nu_{_{T}}=u_0 \, \ell_0/3$ is the turbulent (eddy) viscosity, $u_0$ is the turbulent velocity at the integral turbulent scale $\ell_0$ and
$N^2 = \beta \, |\nabla_z \meanT_{0}|$.

In Eqs.~(\ref{B1})--(\ref{B2}) we use the following dimensionless parameters:
\begin{itemize}
\item{
the effective Rayleigh number:
\begin{eqnarray}
{\rm Ra}_{_{T}} = {L_z^4 \, N^2 \over \nu_{_{T}} \, \kappa_{_{T}}} ,
\label{B3}
\end{eqnarray}
}
\item{
the turbulent Prandtl number:
\begin{eqnarray}
{\rm Pr}_{_{T}} = {\nu_{_{T}} \over \kappa_{_{T}}} ,
\label{B4}
\end{eqnarray}
}
\item{
the scale separation parameter:
\begin{eqnarray}
\epsilon = {\ell_{0}^2 \over 3 L_z^2} ,
\label{B5}
\end{eqnarray}
}
\item{
the non-dimensional total vertical heat flux:
\begin{eqnarray}
\Phi_c = {3 \over \epsilon^2\, {\rm Ra}_{_{T}}} \, \left({u_c \over u_{0}}\right)^3  = {\sigma \over \epsilon^2\, {\rm Ra}_{_{T}}},
\label{B6}
\end{eqnarray}
}
\end{itemize}
where $\sigma = 3 \, (u_c/ u_{0})^3$, $u_c = (\beta F_z^\ast \ell_{0})^{1/3}$ is
the characteristic turbulent convective velocity,
$F_z^\ast$ is the vertical background turbulent flux of potential temperature and $\kappa_{_{T}}$ is the turbulent (eddy) diffusivity.
In Eqs.~(\ref{B1})--(\ref{B2}) we have neglected small terms $\sim{\rm O}(\epsilon^2)$,
where $\epsilon \ll 1$.

Let us discuss the large-scale convective-wind instability, using the stress-free boundary conditions for the velocity field in the vertical direction (along the $z$ axis):
\begin{eqnarray}
\tilde U_z(t, z=0) = \tilde U_z(t, z=1)=0 ,
\label{B40}
\end{eqnarray}
\begin{eqnarray}
\nabla_z \tilde U_x(t, z=0) = \nabla_z \tilde U_x(t, z=1)=0 ,
\label{BBB41}
\end{eqnarray}
\begin{eqnarray}
\nabla_z \tilde U_y(t, z=0) = \nabla_z \tilde U_y(t, z=1)=0 .
\label{B41}
\end{eqnarray}

For a classical laminar convection with the boundary conditions
given by Eqs.~(\ref{B40})--(\ref{B41}), the critical
Rayleigh number required for the onset of convection
is ${\rm Ra}_{\rm cr} \approx 657.5$ \cite{EGKR06,CH61,DR02,RH58}.
The classical laminar convection is described by Eqs.~(\ref{B1})--(\ref{B2}),
where there are no terms $\propto \epsilon$ in Eq.~(\ref{B2}), and
turbulent viscosity and turbulent heat conductivity
are replaced by the molecular viscosity and heat conductivity.

Now let us consider the linear two-dimensional problem
using Eqs.~(\ref{B1})--(\ref{B2}) with $\epsilon \not=0$,
where we take into account the modification of the turbulent heat flux
by nonuniform large-scale motions.
A particular solution of the linearised Eqs.~(\ref{B1})--(\ref{B2}) satisfying the vertical boundary conditions~(\ref{B40})--(\ref{B41}), is given by
\begin{eqnarray}
&& \tilde U_z(t, y, z) = U_0 \sin(\pi z) \cos(\alpha \pi y) \,
\exp(\gamma_{\rm inst} \, t),
\label{B21}
\end{eqnarray}
\begin{eqnarray}
&& \tilde U_y(t, y, z) = - \alpha^{-1} \, U_0 \cos(\pi z) \sin(\alpha \pi y) \,
\exp(\gamma_{\rm inst} \, t),
\nonumber\\
\label{B23}
\end{eqnarray}
\begin{eqnarray}
&& \tilde \Theta(t, y, z) = \Theta_0 \sin(\pi z) \cos(\alpha \pi y) \,
\exp(\gamma_{\rm inst} \, t),
\label{B22}
\end{eqnarray}
and $\tilde U_x(t, y, z)=0$, where $\gamma_{\rm inst}$ is the non-dimensional growth rate of the convective-wind instability, and the parameter $\alpha$ to be determined below.
Equations~(\ref{B21})--(\ref{B22}) imply that the nondimensional wavenumbers
$K_y= \alpha \pi$ and $K_z= \pi$, so that $K=(K_y^2 + K_z^2)^{1/2}= \pi \, (1 + \alpha^2)^{1/2}$.
Note that for a classical laminar convection (where there are no terms $\propto \epsilon$ in Eq.~(\ref{B2}), and $\alpha = 1 /\sqrt{2}$) with the vertical boundary conditions given by Eqs.~(\ref{B40})--(\ref{B41}), the maximum growth rate
of the convective instability is attained at the nondimensional wavenumber $K_y=\pi /\sqrt{2}$.

Let us consider convective turbulence,
and the large-scale properties of the system are described
by the mean-field equations~(\ref{B1}) and~(\ref{B2}), where $\epsilon \not=0$.
Substituting solution (\ref{B21})--(\ref{B22})
into linearised Eqs.~(\ref{B1})--(\ref{B2}), we obtain
the nondimensional growth rate $\gamma_{\rm inst}$ of the large-scale convective-wind instability (see Appendix~\ref{Appendix:B}):
\begin{eqnarray}
\gamma_{\rm inst} &=& {\alpha \over (1+ \alpha^2)^{1/2}} \left[{\rm Ra}_{_{T}} + {\sigma \, \pi^2 \over 2 \epsilon}
\left(1 - \alpha^2\right) \right]^{1/2}
\nonumber\\
&&- \pi^2 \, (1+ \alpha^2) .
\label{C1}
\end{eqnarray}
Here we consider the case when the turbulent Prandtl number
${\rm Pr}_{_{T}}=1$.
The condition $\gamma_{\rm inst}=0$  yields the critical effective Rayleigh number ${\rm Ra}_{_{T}}^{\rm cr}$
required for the excitation of the large-scale convective-wind instability:
\begin{eqnarray}
{\rm Ra}_{_{T}}^{\rm cr} &=& {\pi^4 \, (1+ \alpha^2)^{3} \over \alpha^2}
\, \left[1 - {\sigma \,(1 - \alpha^2) \over 2 \pi^2 \, \epsilon \, (1+ \alpha^2)^{3}}
\right] .
\label{RC1}
\end{eqnarray}
Note that Eq.~(\ref{RC1}) is valid for arbitrary turbulent Prandtl number
${\rm Pr}_{_{T}}$.

Equations~(\ref{C1}) and~(\ref{RC1}) at $\epsilon=0$ (i.e., $\sigma=0$) describe the classical laminar convection,
where the effective Rayleigh number ${\rm Ra}_{_{T}}$ based on the turbulent transport coefficients
should be replaced by the Rayleigh number ${\rm Ra}$ based on the molecular transport coefficients.
It follows from Eq.~(\ref{RC1}), that the critical effective Rayleigh number
required for the large-scale convective-wind instability is strongly reduced
in turbulence (for perturbations with $\alpha<1$).
This is due to the modification of the turbulent heat flux
by the nonuniform motions.
This effect will be discussed in the next section
where the results of mean-field simulations are described.
Note that Eq.~(\ref{C1}) agrees with Eq.~(\ref{R200}) for ${\rm Ra}_{_{T}} \to 0$.

\begin{figure}
\vspace*{1mm}
\centering
\includegraphics[width=6.5cm]{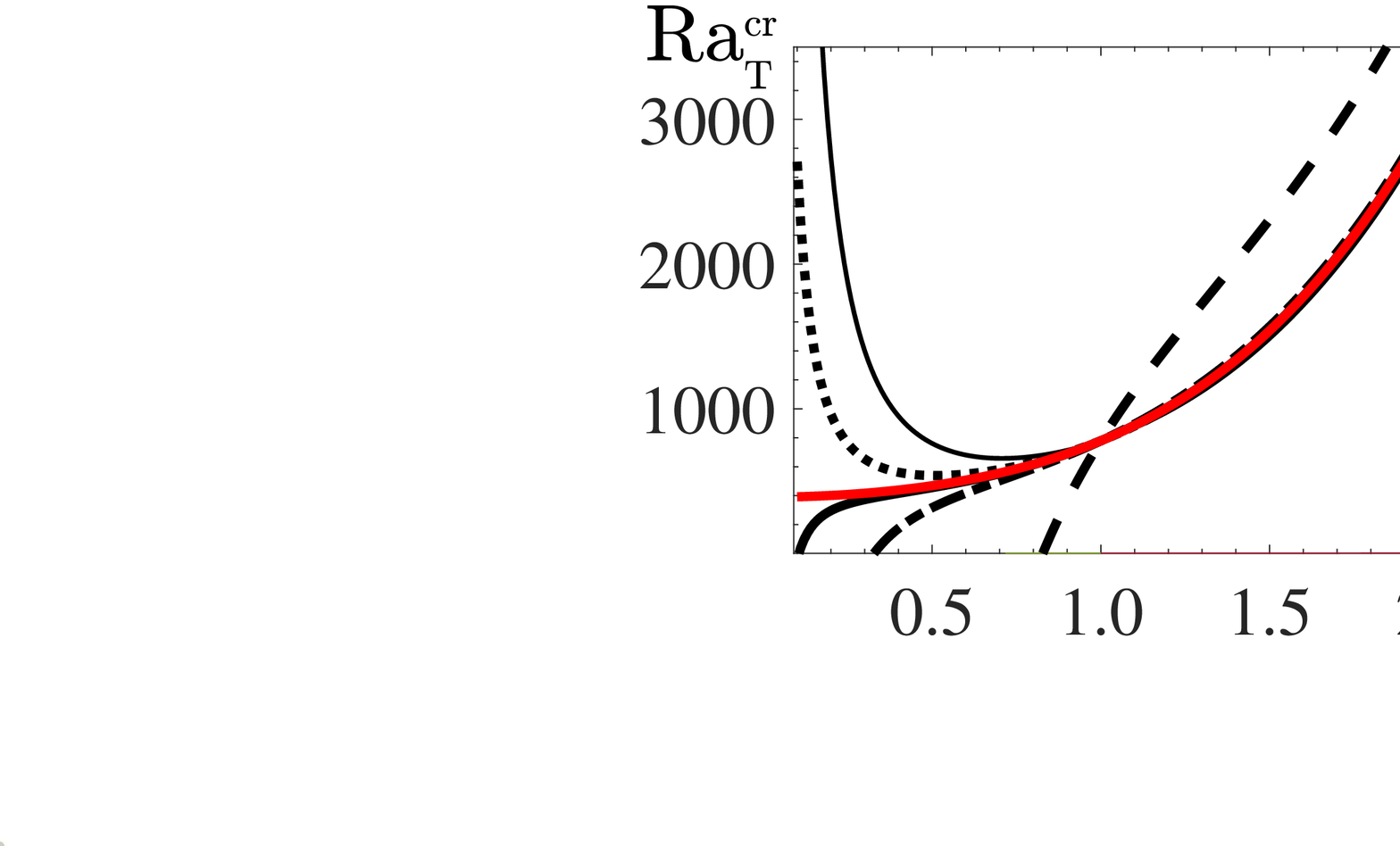}
\caption{\label{Fig2} The critical effective Rayleigh number
${\rm Ra}_{_{T}}^{\rm cr}$ versus the parameter $\alpha$ for $\epsilon =10^{-2}$
and different values of the parameter $\sigma=$ 3 (dashed); 0.3 (dashed-dotted) 0.206 (solid thick); 0.197 (red); 0.15 (dotted); and for the classical laminar convection with $\sigma =0$ (solid thin).}
\end{figure}

\begin{figure}
\vspace*{1mm}
\centering
\includegraphics[width=7cm]{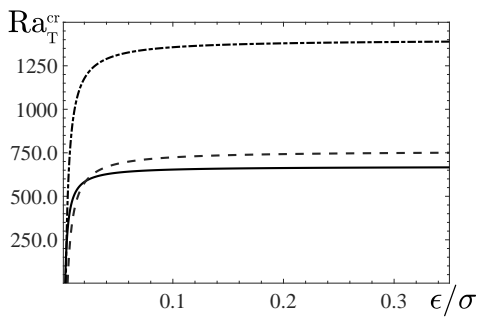}
\caption{\label{Fig3}
The critical effective Rayleigh number
${\rm Ra}_{_{T}}^{\rm cr}$ versus the parameter $\epsilon/\sigma$
and different values of the parameter $\alpha=$ 0.3 (dashed-dotted); 0.5 (dashed) and 0.8 (solid).
}
\end{figure}

\begin{figure}
\vspace*{1mm}
\centering
\includegraphics[width=7cm]{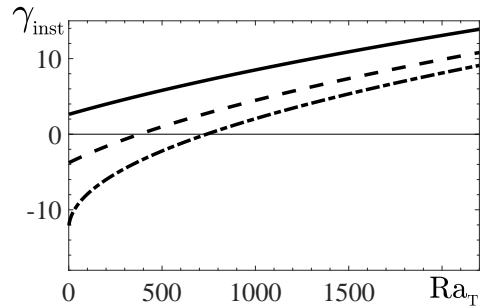}
\caption{\label{Fig4} The non-dimensional growth rate $\gamma_{\rm inst}$ of the large-scale convective-wind
instability versus the effective Rayleigh numbers ${\rm Ra}_{_{T}}$
for $\alpha=0.5$, $\epsilon =10^{-2}$ and different values of the parameter $\sigma=$ 3 (solid);
1 (dashed); and for the classical convection with $\sigma =0$ (dashed-dotted).
}
\end{figure}

\begin{figure}
\vspace*{1mm}
\centering
\includegraphics[width=6.5cm]{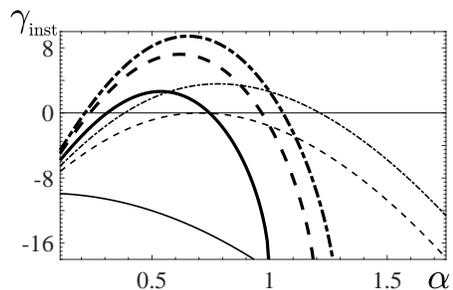}
\caption{\label{Fig5} The non-dimensional growth rate $\gamma_{\rm inst}$ of the large-scale convective-wind instability versus $\alpha$ for $\sigma=3$, $\epsilon =10^{-2}$ and different effective Rayleigh numbers ${\rm Ra}_{_{T}}=$ 1000 (dashed-dotted thick); 657.5 (dashed thick); 0.5 (solid-thick); and for the classical convection with $\sigma =0$ and different Rayleigh number ${\rm Ra}_{_{T}}=$ 1000 (dashed-dotted thin); 657.5 (dashed thin); 0.5 (solid-thin).
}
\end{figure}

In Fig.~\ref{Fig2} we show the critical effective Rayleigh number
${\rm Ra}_{_{T}}^{\rm cr}$ versus the parameter $\alpha$ for
different values of the parameter $\sigma= 3 \, (u_c/ u_{0})^3$,
while in Fig.~\ref{Fig3} we plot the critical effective Rayleigh number
${\rm Ra}_{_{T}}^{\rm cr}$ as the function of the parameter $\epsilon/\sigma$
for different values of the parameter $\alpha$.
The classical convection corresponds to $\sigma=0$, when the turbulent integral scale
$\ell_{0}$ vanishes.
The increase of the parameter $\sigma$ decreases
the critical effective Rayleigh number required for the large-scale convective-wind instability.
This tendency is also seen in Fig.~\ref{Fig4}, where we plot
the non-dimensional growth rate $\gamma_{\rm inst}$ of the large-scale convective-wind instability versus the effective Rayleigh numbers ${\rm Ra}_{_{T}}$ for different values of the parameter $\sigma$.
In Fig.~\ref{Fig5} we show the non-dimensional growth rate $\gamma_{\rm inst}$ of the large-scale convective-wind instability versus $\alpha$ for different effective Rayleigh numbers ${\rm Ra}_{_{T}}$.
The thick lines in Fig.~\ref{Fig5} correspond to turbulent convection,
while the thin lines correspond to the classical laminar convection.
In the classical laminar convection the maximum growth rate of the convective instability
is attained when $\alpha$ tends to 1, while in the turbulent convection
the convective-wind instability is more effective when $\alpha$ tends to 1/2.

\begin{figure}
\vspace*{1mm}
\centering
\includegraphics[width=7cm]{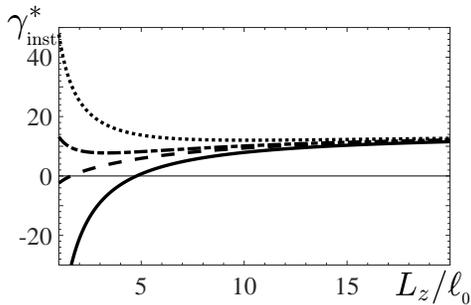}
\caption{\label{Fig6}
The non-dimensional growth rate $\gamma_{\rm inst}^\ast = 10 \, \gamma_{\rm inst} \, \epsilon^{1/2}$ of the large-scale convective-wind instability versus $L_z/\ell_0$ for $\alpha =0.5$ and different effective Rayleigh number ${\rm Ra}_{_{T}}=$ 2000 (dotted); 1000 (dashed-dotted); 657.5 (dashed); 0.5 (solid).
}
\end{figure}

In Fig.~\ref{Fig6} we plot the nondimensional growth rate $\gamma_{\rm inst}^\ast = 10 \, \gamma_{\rm inst} \, \epsilon^{1/2}$ of the large-scale convective-wind instability versus $L_z/\ell_0$ for different values of the effective Rayleigh number.
We show the growth rate $\gamma_{\rm inst}^\ast$ instead of $\gamma_{\rm inst}$
by the following reasons.
Since the growth rate of the large-scale convective-wind instability is
measured in the units of $\nu_{_{T}}/L_z^2$ and the turbulent viscosity $\nu_{_{T}}=u_0 \, \ell_0/3$
is proportional to the integral turbulence scale $\ell_0$, we show
the nondimensional growth rate $\gamma_{\rm inst}^\ast$ which is measured in the units
which are independent of $\ell_0$.
Figure~\ref{Fig6} demonstrates that when the effective Rayleigh number is not large,
the separation of scales $L_z/\ell_0$ between the size of the convective layer and the integral scale
of turbulence should be large (more than 5 - 10) to observe the large-scale convective-wind instability
and formation of large-scale coherent structures.

\section{Results of the mean-field numerical simulations}
\label{sec-IV}

In this section we study nonlinear evolution of the convective-wind instability
by means of the mean-field numerical simulations, where
we solve numerically Eqs.~(\ref{B1})--(\ref{B2}) for
the periodic boundary conditions in the horizontal directions
(along the $x$ and $y$ axes).
We use two kinds of the vertical boundary conditions
for the velocity field:  (i) stress-free boundary conditions
corresponding to the free vertical boundaries and (ii) no-slip boundary conditions
corresponding to the rigid vertical boundaries.
The boundary conditions for the potential temperature in the vertical direction are
$\tilde \Theta(t, z=0) = \tilde \Theta(t, z=1)=0$.

Simulations are performed using the ANSYS FLUENT code (version 19.2)
[https://lawn-mower-manual.com/htm/ansys-fluent-theory-guide-2020]
which applies the Final Volume method
in the $3D$ box $L_x=L_z=1$ and $L_y=2$.
The additional terms $\propto \epsilon$ in the divergence of the turbulent heat flux are implemented
into the code as the source terms in the potential temperature equation~(\ref{B2}).

The simulations are performed with the spatial resolution
$100 \times 200 \times 100$  in $x$, $y$ and $z$ directions, respectively.
A sensitivity check has been also made for the spatial resolution
$150  \times 300 \times 150$.
For both cases similar results for velocity and potential temperature
have been obtained.
In all simulations, we use a time step of $10^{-3}$.
A sensitivity check has been made for time steps as well.
Time steps of $10^{-3}$ and $2 \times 10^{-3}$ have been tested
and a maximum error of 0.05 \%
at velocity and potential temperature using these time steps has been obtained.
In addition, a convergence error was set to be less than $10^{-6}$.

The parameters for the simulations are the following: the turbulent Prandtl number
${\rm Pr}_{_{T}}=1$, the ratio $u_c / u_{0}=1$, the effective Rayleigh number varies from
${\rm Ra}_{_{T}}=0.5$ to ${\rm Ra}_{_{T}}=2700$
and the scale separation parameter $\epsilon$ varies in the range $(0.5 - 2) \times 10^{-3}$.
Different theoretical and numerical studies show that the turbulent Prandtl number ${\rm Pr}_{_{T}}$
for large Reynolds numbers in isothermal and convective turbulence is of the order of 1
(see Refs.~\cite{MC90,AA99,YII03,EKR96b,JJ16}), while
in stably stratified turbulence the turbulent Prandtl number can be much
larger than 1 only for large Richardson numbers  (see Refs.~\cite{KA94,L19,SR10,O01,ZKR07,ZKR13,KRZ21}).
The variations of the parameter $\epsilon=(0.5 - 2.5) \times 10^{-3}$
(characterizing the separation of scales between
the height $L_z$ of the convective layer and the integral turbulence scale $\ell_0$)
corresponds to variations of the ratio $L_z/\ell_0= 12 - 26$.
The choice of the other parameters is discussed below and in Sect.~\ref{sec-V}.

\subsection{Numerical simulations with the stress-free vertical boundary conditions}
\label{subsec-IV-A}

Let us discuss the results of the mean-field numerical simulations
which take into account the modification of the turbulent heat flux by the non-uniform motions
[described by the terms $\propto \epsilon$ in Eq.~(\ref{B2})].
Here we discuss the numerical results for the stress-free
boundary conditions in the vertical direction and for the following initial conditions:
\begin{eqnarray}
&& \tilde U_z(t=0, z) = U_0 \sin(\pi z) \cos(\alpha \pi y) ,
\label{BB21}
\end{eqnarray}
\begin{eqnarray}
&& \tilde U_y(t=0, z) = - \alpha^{-1} \, U_0 \cos(\pi z) \sin(\alpha \pi y) ,
\label{BB23}
\end{eqnarray}
\begin{eqnarray}
&& \tilde U_x(t=0, z) = 0 ,
\end{eqnarray}
\begin{eqnarray}
&& \tilde \Theta(t=0, z) = \Theta_0 \sin(\pi z) \cos(\alpha \pi y) ,
\label{BB22}
\end{eqnarray}
[see analytical solution~(\ref{B21})--(\ref{B22})],
where $U_0$  is initial amplitude of the nondimensional mean vertical velocity.
In the mean-field numerical simulations, we use the parameters $U_0=10^{-4}$ and $\alpha=0.35$.
In some simulations we also use $U_0=0.23$ to avoid a long transition range.
The obtained results are nearly independent of parameter $U_0$.

\begin{figure}
\vspace*{1mm}
\centering
\includegraphics[width=8cm]{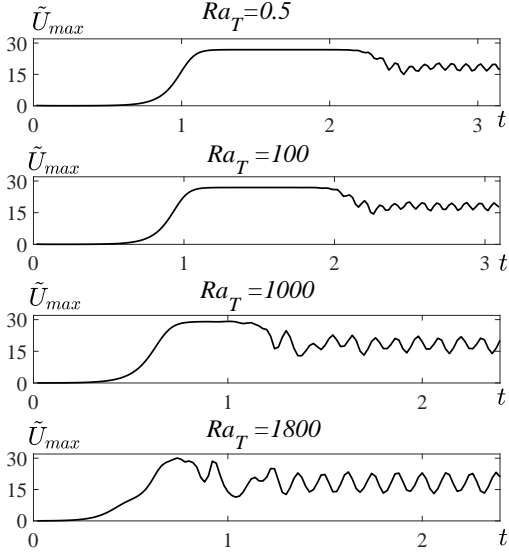}
\caption{\label{Fig7}
Time evolution of the maximum velocity $\tilde {U}_{\rm max}(t)$ at $\epsilon=2.5 \times 10^{-3}$
and different values of the effective Rayleigh number ${\rm Ra}_{_{T}}$
for the stress-free boundary conditions.
}
\end{figure}

\begin{figure}
\vspace*{1mm}
\centering
\includegraphics[width=8cm]{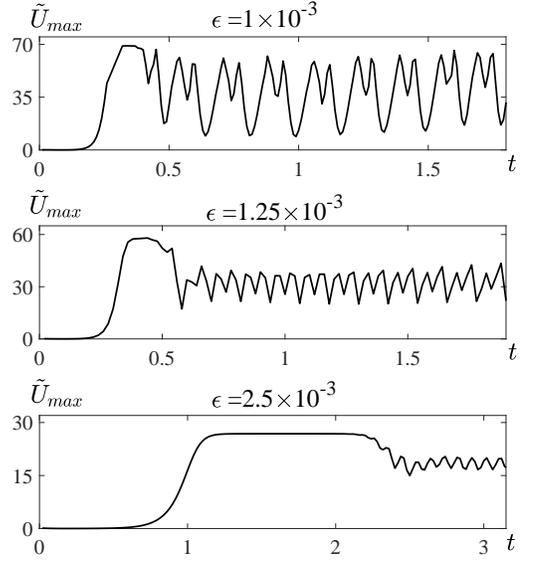}
\caption{\label{Fig8}
Time evolution of the maximum velocity $\tilde {U}_{\rm max}(t)$ at ${\rm Ra}_{_{T}}=0.5$
and different values of $\epsilon$ for the stress-free boundary conditions.
}
\end{figure}

In Figs.~\ref{Fig7}--\ref{Fig8} we show the time evolution of the maximum velocity $\tilde {U}_{\rm max}(t)$
for different effective Rayleigh numbers varying from  ${\rm Ra}_{_{T}}=0.5$
to $1800$ and different values of the parameter $\epsilon$ (which characterizes
scale separation between the vertical size $L_z$ of the computational domain and the integral turbulence scale $\ell_0$).
During the linear stage of the large-scale convective-wind instability, the maximum velocity $\tilde {U}_{\rm max}(t)$
grows in time exponentially. The instability  is saturated by the nonlinear effects.
For smaller values of the parameter $\epsilon$ (i.e., for larger values of the parameter
$L_z/\ell_0$), after the stationary stage when the maximum velocity $\tilde {U}_{\rm max}(t)$ reaches the constant,
nonlinear oscillations of the maximum velocity $\tilde {U}_{\rm max}(t)$ are observed.
With the increase of the  effective Rayleigh number ${\rm Ra}_{_{T}}$, the duration of the stationary stage decreases  (see Fig.~\ref{Fig7}).
The values of the maximum velocity $\tilde {U}_{\rm max}$ at the stationary stage are nearly independent of
the  effective Rayleigh number ${\rm Ra}_{_{T}}$ (see Fig.~\ref{Fig7}), but
$\tilde {U}_{\rm max}$ at the stationary stage
strongly depends on the scale separation parameter $\epsilon$ (see Fig.~\ref{Fig8}).

\begin{figure}
\vspace*{1mm}
\centering
\includegraphics[width=8cm]{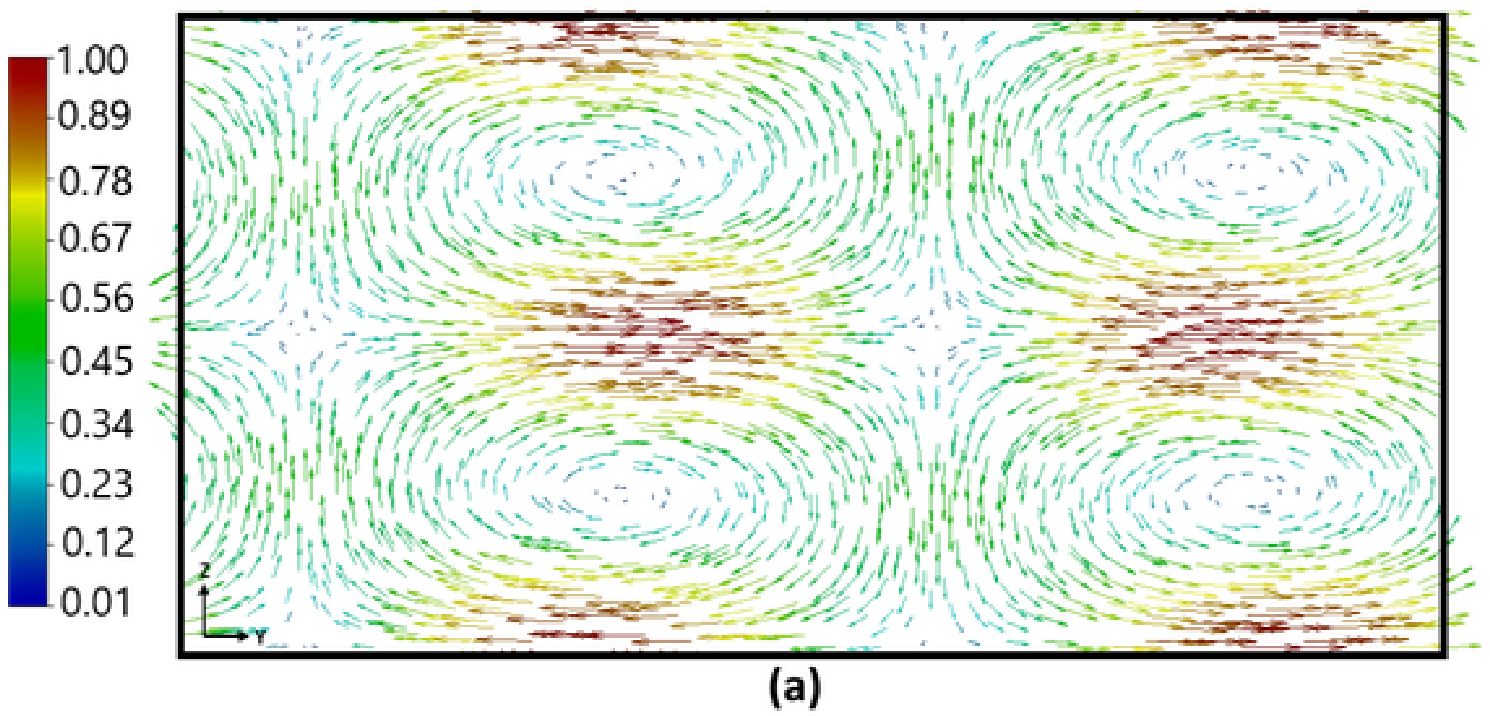}
\includegraphics[width=8cm]{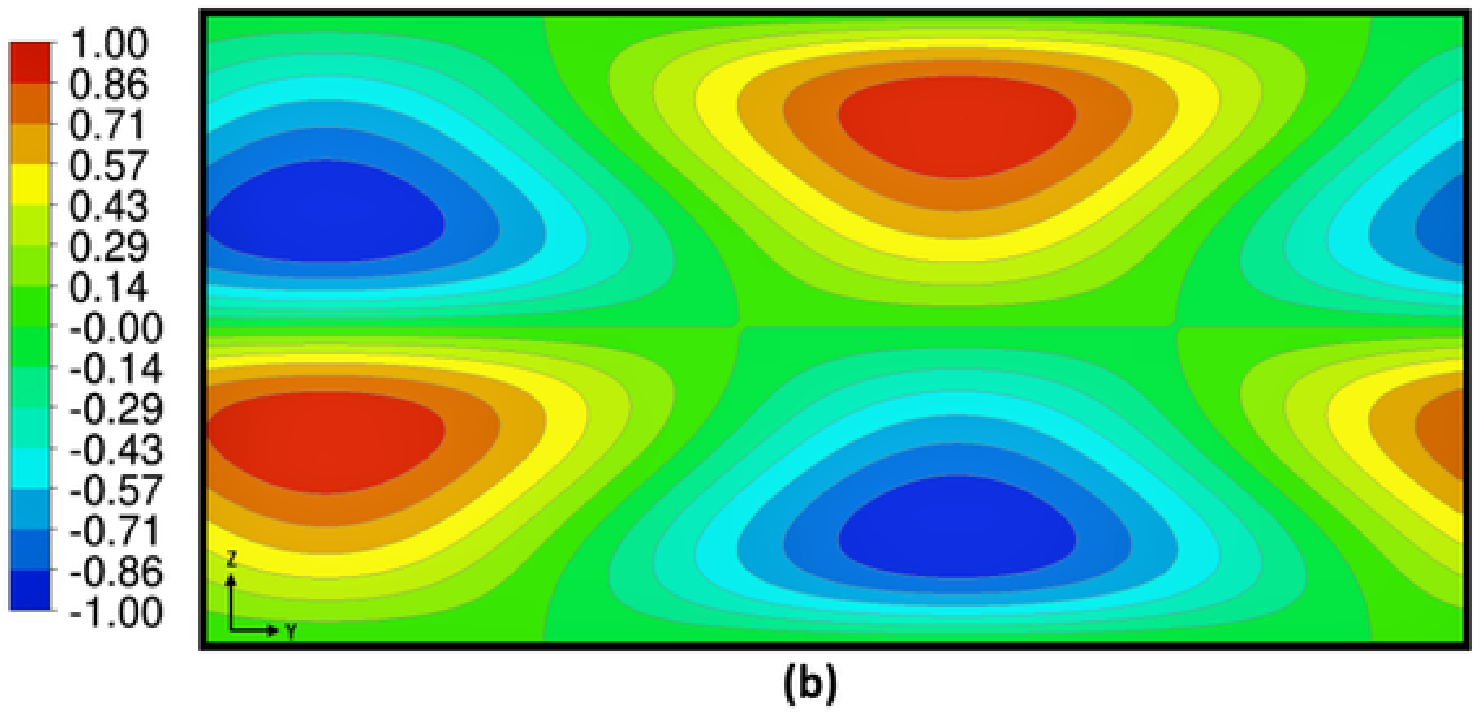}
\includegraphics[width=8cm]{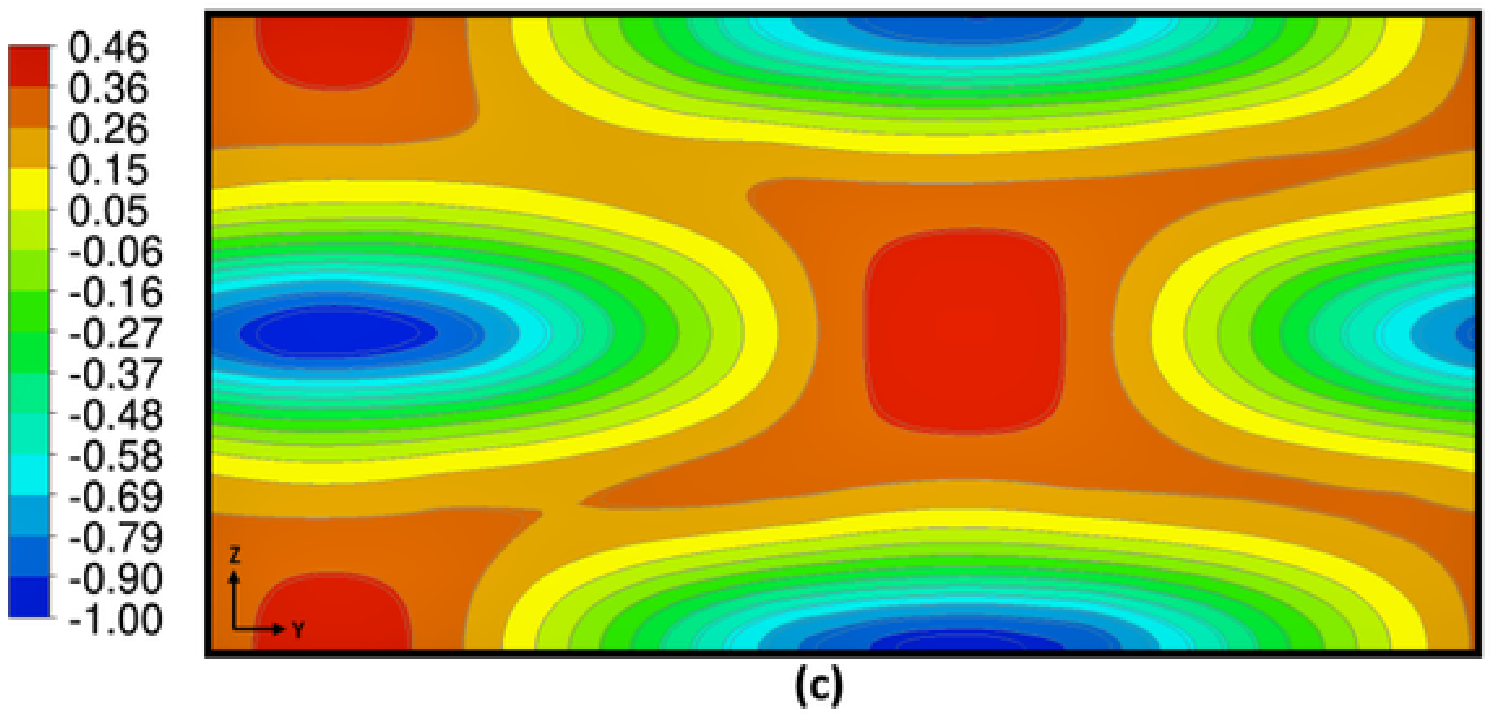}
\caption{\label{Fig9}
The velocity patterns at the stationary stage (a),
the potential temperature deviations $\tilde \Theta \, {\rm Ra}_{_{T}}$ from the equilibrium potential
temperature in the basic reference state (b) and
the total vertical gradient of the mean potential temperature $(\nabla_z \tilde \Theta -1) \, {\rm Ra}_{_{T}}$ (c)
in $yz$ plane for the stress-free vertical boundary conditions for $\epsilon=2.5 \times 10^{-3}$
and the effective Rayleigh number ${\rm Ra}_{_{T}}=0.5$.
All fields are normalized by their maximum values.
}
\end{figure}

We show also the velocity patterns in the $yz$ plane at the stationary stage (Fig.~\ref{Fig9}a).
Since the potential temperature is measured in the units of $L_z \, N^2 \, {\rm Pr}_{_{T}}/\beta
= {\rm Ra}_{_{T}} \nu_{_{T}}^2/(\beta L_z^3)$,
we show in Fig.~\ref{Fig9}b the pattern of the
normalized deviations of the potential temperature $\tilde \Theta \, {\rm Ra}_{_{T}}$ from
the equilibrium potential temperature in the basic reference state.
Figure~\ref{Fig9} is for the stress-free vertical boundary conditions, $\epsilon=2.5 \times 10^{-3}$
and the effective Rayleigh number, ${\rm Ra}_{_{T}}=0.5$.
All fields in Fig.~\ref{Fig9} are normalized by their maximum values.

Figures~\ref{Fig9}a--\ref{Fig9}b demonstrate the four-cell patterns of the velocity and potential temperature, where
the two-cell patterns are located in both, the upper and bottom parts of Fig.~\ref{Fig9}.
Remarkably, the large-scale circulations exist even below the threshold of the laminar convection.
The main reason is that turbulence with nonuniform large-scale flows contributes to the
turbulent heat flux. In particular, nonuniform large-scale flows produce anisotropic
velocity fluctuations modifying the turbulent heat flux.
As the result, the evolutionary equation~(\ref{B2}) for the potential temperature $\tilde \Theta$
contains the new terms proportional to the spatial derivatives of the mean velocity field $\tilde {\bm U}$ (see the terms $\propto \epsilon$).
For the stress-free boundary conditions and different effective Rayleigh numbers varying from ${\rm Ra}_{_{T}}=0.5$ to ${\rm Ra}_{_{T}}=1800$, we observe the same four-cell flow patterns which is seen in Fig.~\ref{Fig9}.

Since the total gradient of the potential temperature is the sum of the equilibrium constant
gradient of the potential temperature $\nabla_z \meanT_{\rm eq}$ (which is negative as usual for a convection) and the gradient of the potential temperature $\nabla_z \tilde \Theta$, we show in Fig.~\ref{Fig9}c the pattern of the normalized total vertical gradient of the mean potential temperature, $(\nabla_z \tilde \Theta -1) \, {\rm Ra}_{_{T}}$ that
includes the equilibrium constant gradient of the potential temperature $\nabla_z \meanT_{\rm eq}$.
The normalized vertical gradient of the mean potential temperature, $(\nabla_z \tilde \Theta -1) \, {\rm Ra}_{_{T}}$ describes convection. Figure~\ref{Fig9}c demonstrates existence of the regions with the positive gradient of the potential temperature $(\nabla_z \tilde \Theta -1) \, {\rm Ra}_{_{T}}$.
This implies that these regions are stably stratified.
Such effects have been previously observed in experiments \cite{NSS01,BEKR20} and direct numerical simulations \cite{AB16,KRB17,KA19} of turbulent convection.

The existence of the regions with the positive gradient of the potential temperature
inside the large-scale circulation can be explained as follows.
The total vertical heat flux ${\bm F}_z^{\rm tot}$ is the sum
of the mean vertical heat flux $\meanUU_z \, \meanTheta$ of the
large-scale circulation, the vertical turbulent heat flux
${\bm F}_{z}^\ast=- \kappa_{_{T}} {\bm \nabla}_z \meanTheta$
and the new turbulent heat flux ${\bm F}_z^{\rm new} = - \tau_0 \, {\bm F}_{z}^{\ast} \, {\rm div}
\, \meanUU_{\perp}$ (these fluxes are written here in the dimensional form),
i.e.,
\begin{eqnarray}
{\bm F}_z^{\rm tot}=\meanUU_z \, \meanTheta - \kappa_{_{T}} {\bm \nabla}_z \meanTheta
- \tau_0 \, {\bm F}_{z}^{\ast} \, {\rm div}
\, \meanUU_{\perp} .
\label{PK1}
\end{eqnarray}
Note that the last term in Eq.~(\ref{R100}) does not have the vertical component.
Equation~(\ref{PK1}) yields the vertical gradient of the mean potential temperature
$\nabla_z \meanTheta$:
\begin{eqnarray}
\nabla_z \meanTheta = {\meanU_z \, \meanTheta - F_z^{\rm tot} \over \kappa_{_{T}} \,
(1 - \tau_0 \, {\rm div}\, \meanUU_{\perp})}  .
\label{DDD1}
\end{eqnarray}
In the regions inside the large-scale circulation where $\meanU_z \, \meanTheta > F_z^{\rm tot}$,
the vertical gradient $\nabla_z \meanTheta$ is positive, while when $\meanU_z \, \meanTheta < F_z^{\rm tot}$, the vertical gradient $\nabla_z \meanTheta$ is negative. Note that usually $\tau_0 \, |{\rm div}\, \meanUU_{\perp}| < 1$.
This explains the existence of the regions with the positive gradient
of the potential temperature inside the large-scale circulation in a turbulent convection.

\subsection{Numerical simulations with the no-slip vertical boundary conditions}
\label{subsec-IV-B}

\begin{figure}
\vspace*{1mm}
\centering
\includegraphics[width=8cm]{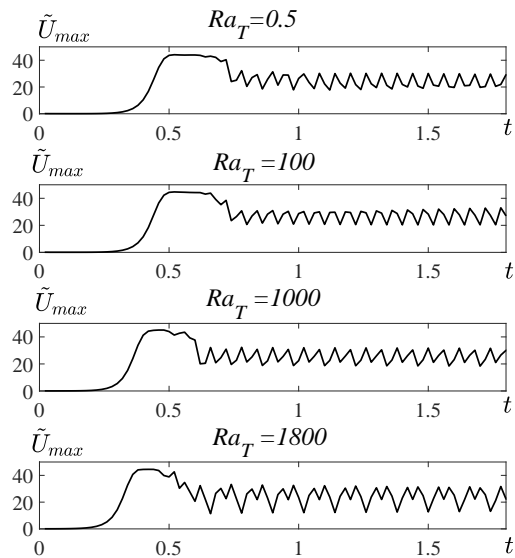}
\caption{\label{Fig10}
Time evolution of the maximum velocity $\tilde {U}_{\rm max}(t)$ at $\epsilon=10^{-3}$
and different values of the effective Rayleigh number ${\rm Ra}_{_{T}}$
for the no-slip boundary conditions.
}
\end{figure}

\begin{figure}
\vspace*{1mm}
\centering
\includegraphics[width=8cm]{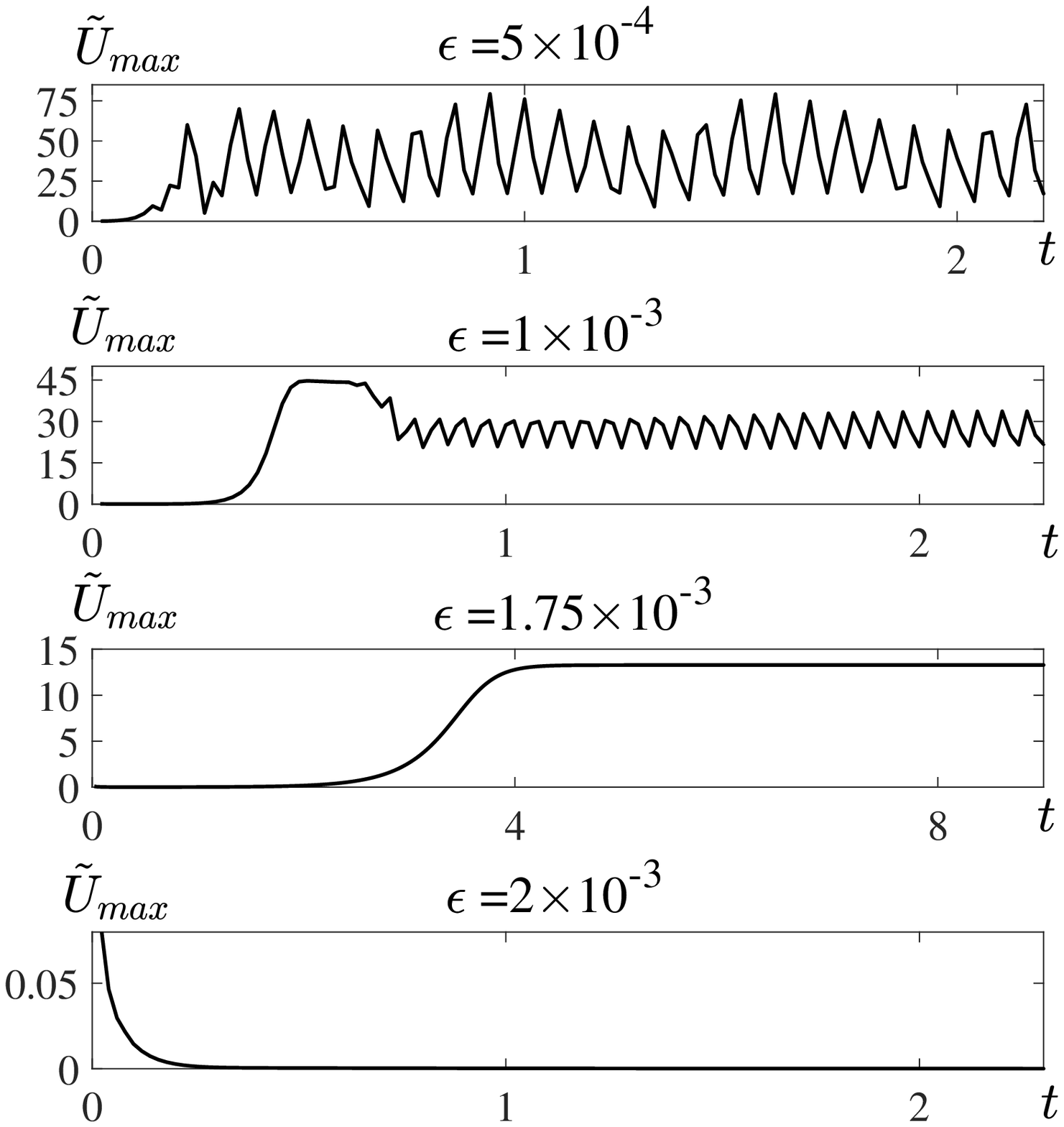}
\caption{\label{Fig11}
Time evolution of the maximum velocity $\tilde {U}_{\rm max}(t)$ at ${\rm Ra}_{_{T}}=100$
and different values of $\epsilon$ for the no-slip boundary conditions.
}
\end{figure}

\begin{figure}
\vspace*{1mm}
\centering
\includegraphics[width=8cm]{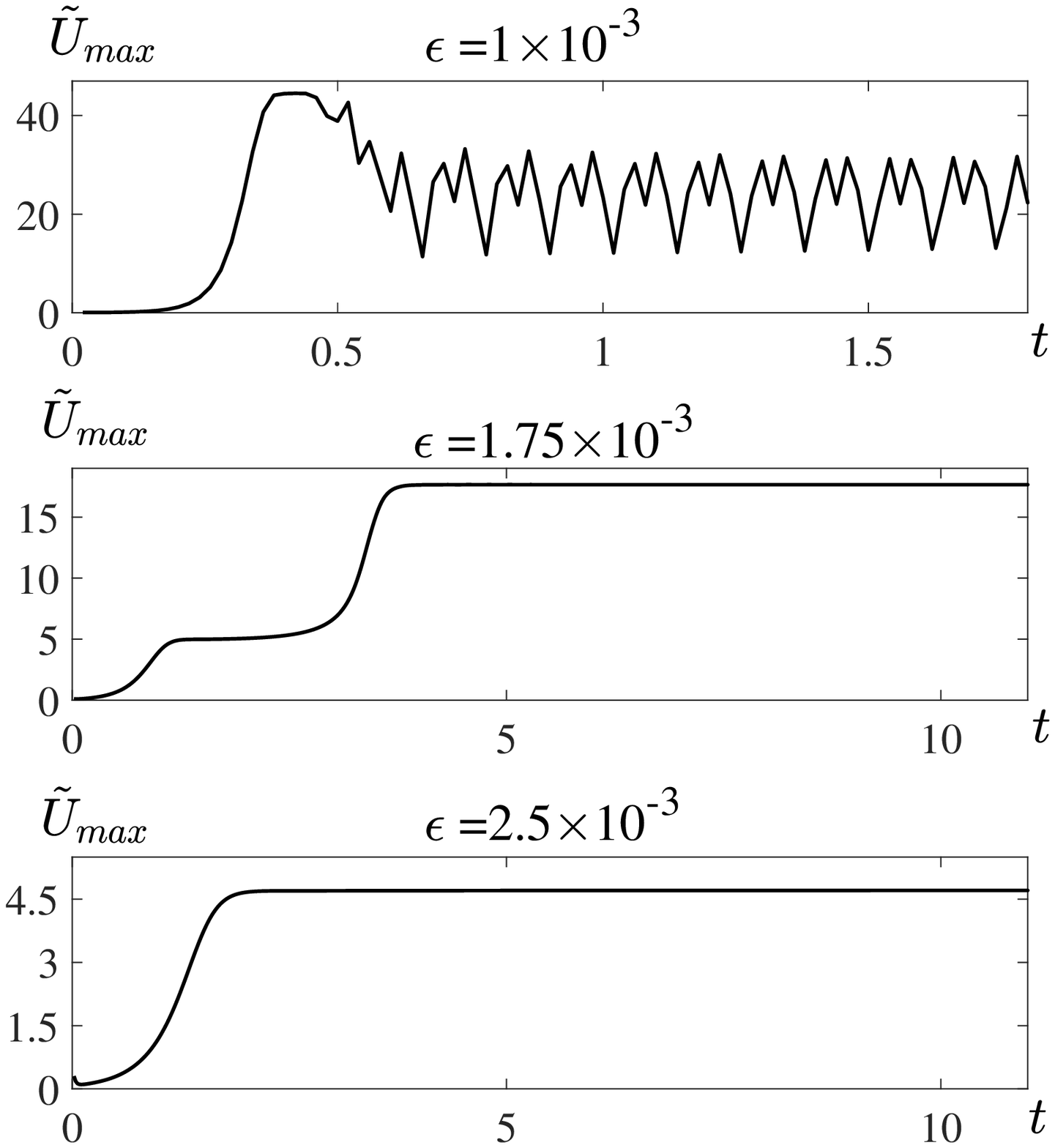}
\caption{\label{Fig12}
Time evolution of the maximum velocity $\tilde {U}_{\rm max}(t)$ at ${\rm Ra}_{_{T}}=1800$
and different values of $\epsilon$ for the no-slip boundary conditions.
}
\end{figure}

\begin{figure}
\vspace*{1mm}
\centering
\includegraphics[width=7.5cm]{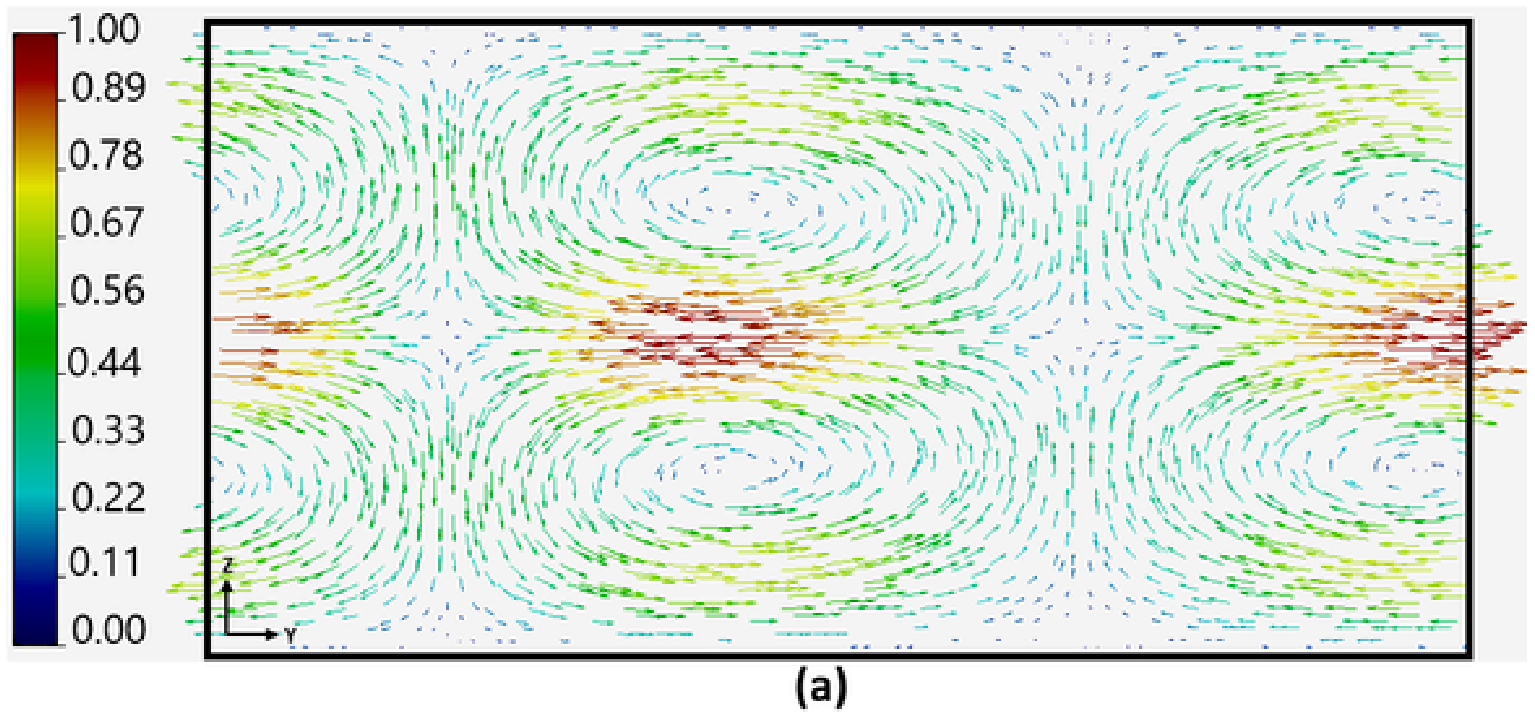}
\includegraphics[width=7.5cm]{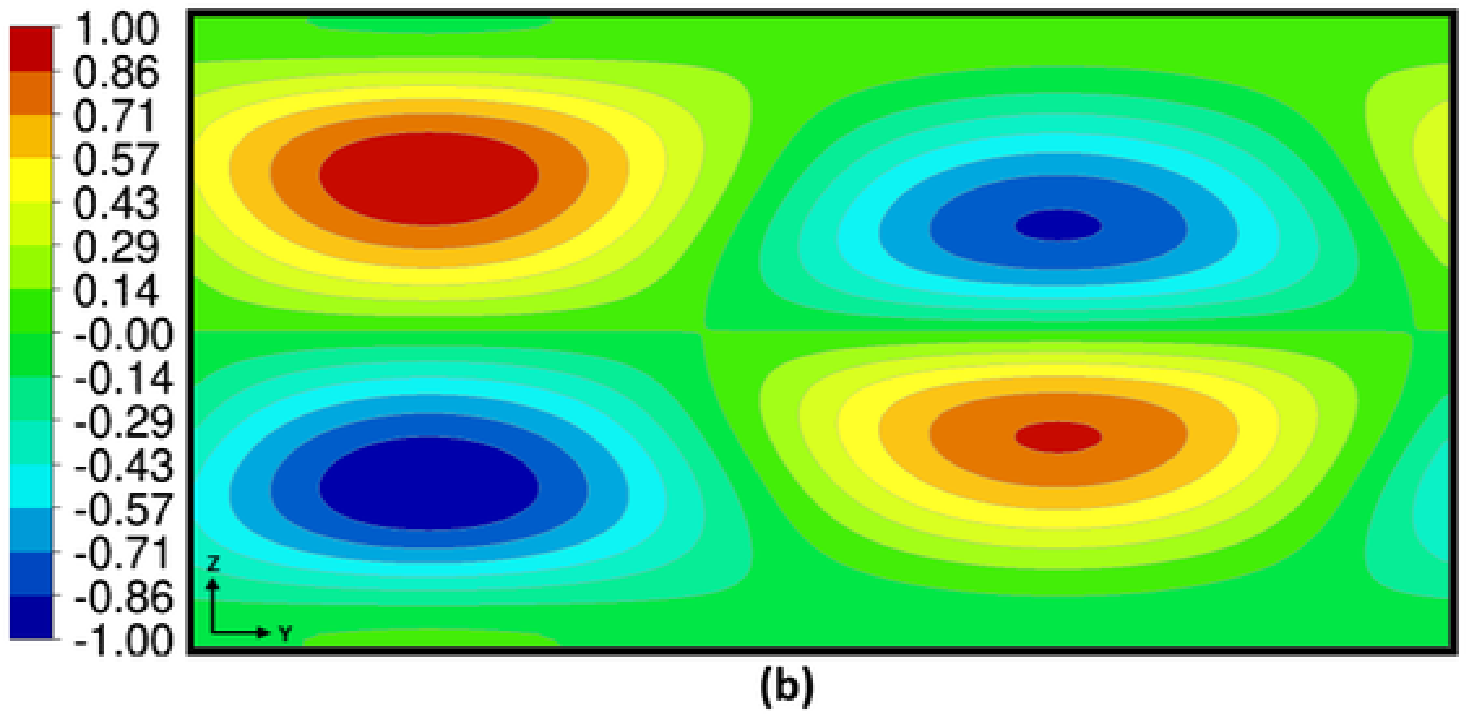}
\includegraphics[width=7.5cm]{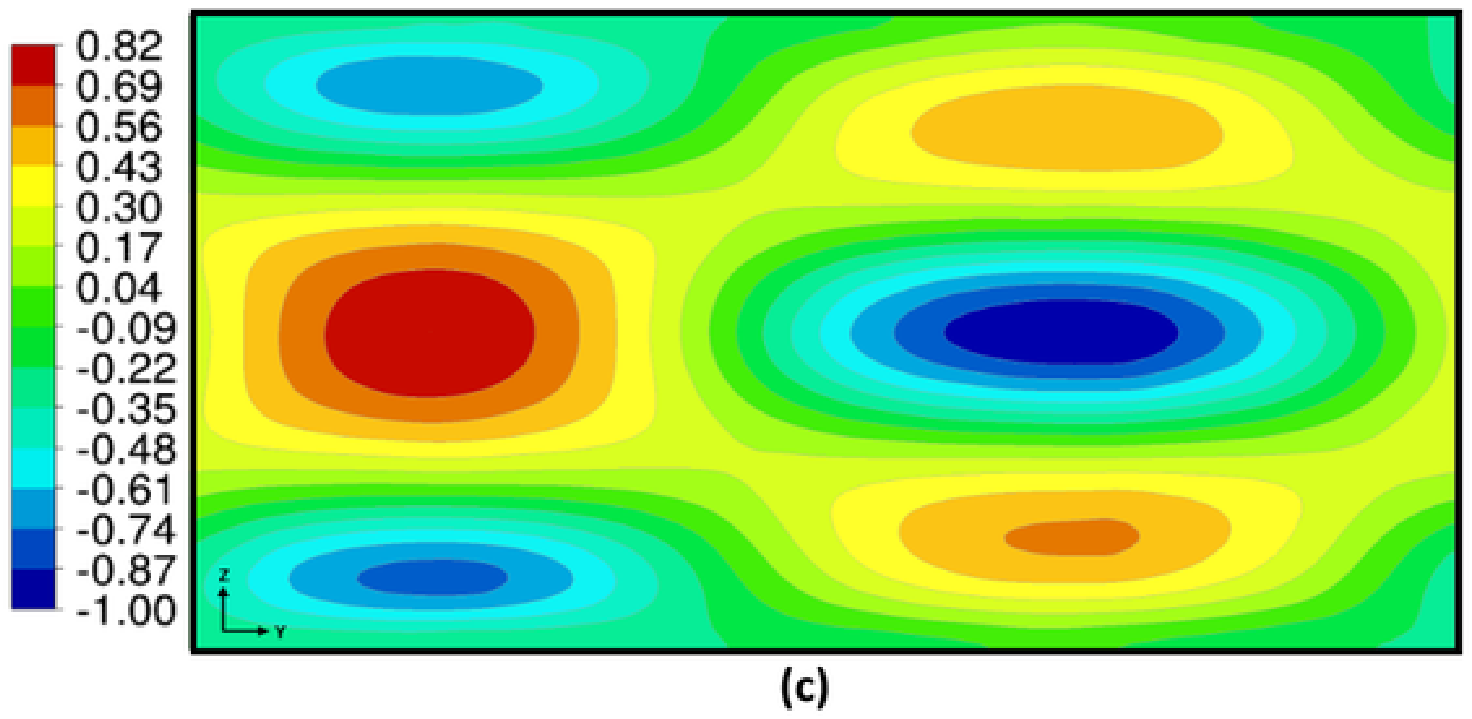}
\caption{\label{Fig13}
The velocity patterns at the stationary stage (a),
the potential temperature deviations $\tilde \Theta \, {\rm Ra}_{_{T}}$ from the equilibrium potential
temperature in the basic reference state (b) and
the total vertical gradient of the mean potential temperature $(\nabla_z \tilde \Theta -1) \, {\rm Ra}_{_{T}}$ (c)
in $yz$ plane for the no-slip vertical boundary conditions for $\epsilon=1.75 \times 10^{-3}$
and the effective Rayleigh number ${\rm Ra}_{_{T}}=1800$.
All fields are normalized by their maximum values.
}
\end{figure}

\begin{figure}
\vspace*{1mm}
\centering
\includegraphics[width=7.5cm]{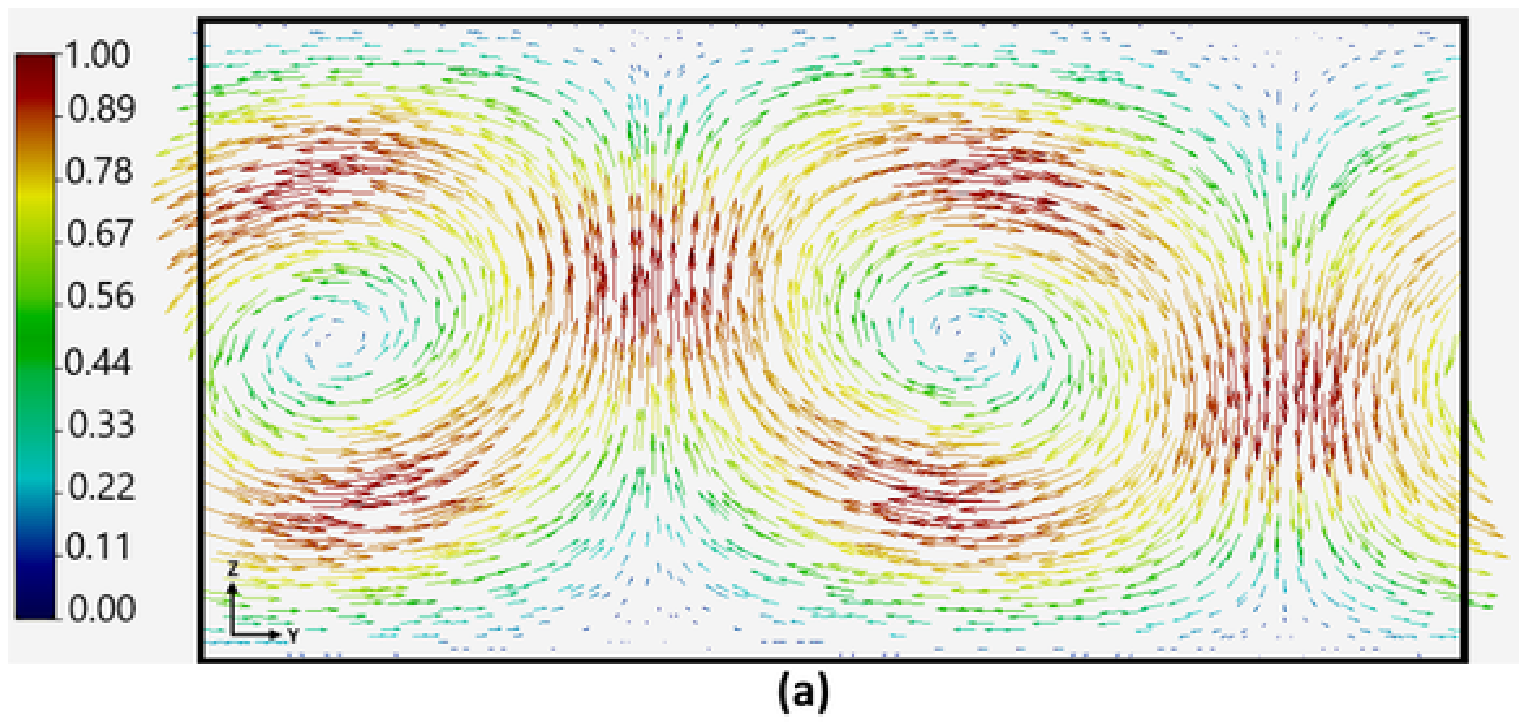}
\includegraphics[width=7.5cm]{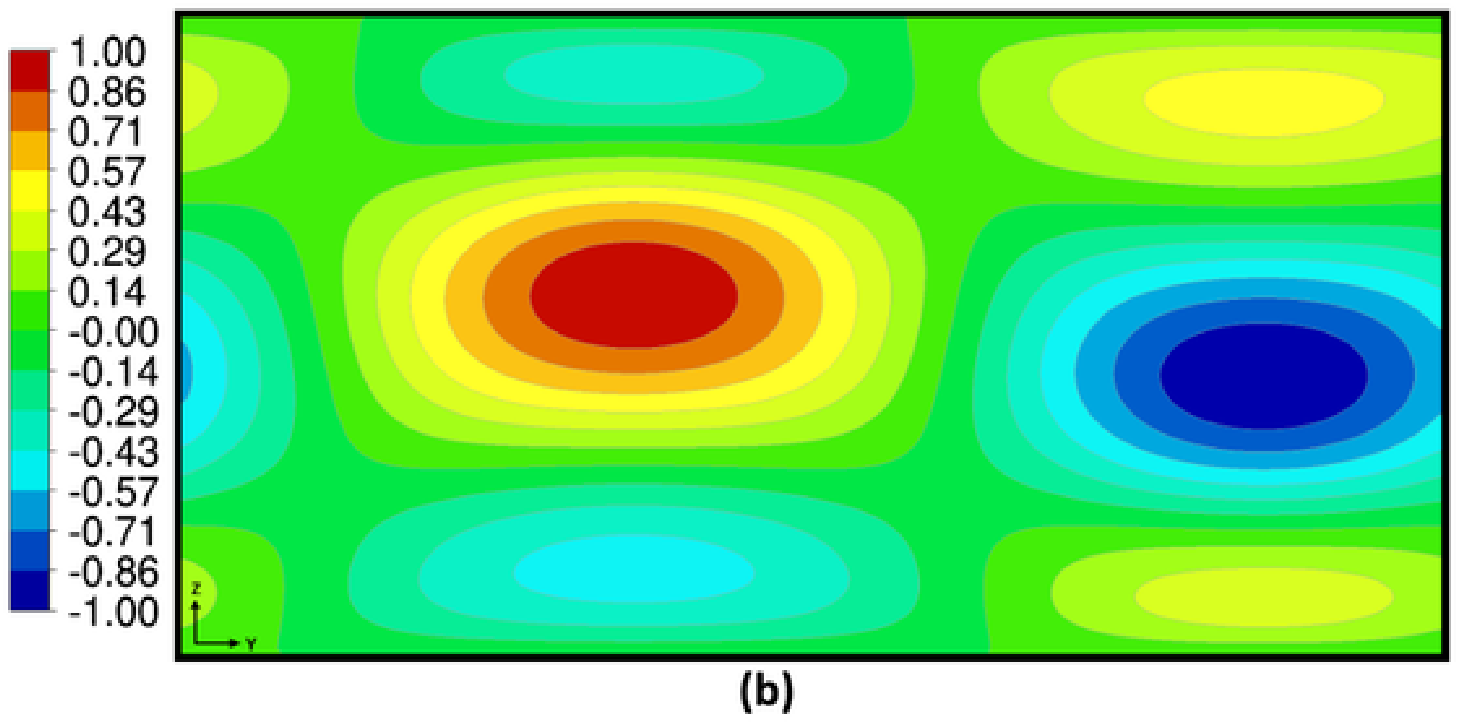}
\includegraphics[width=7.5cm]{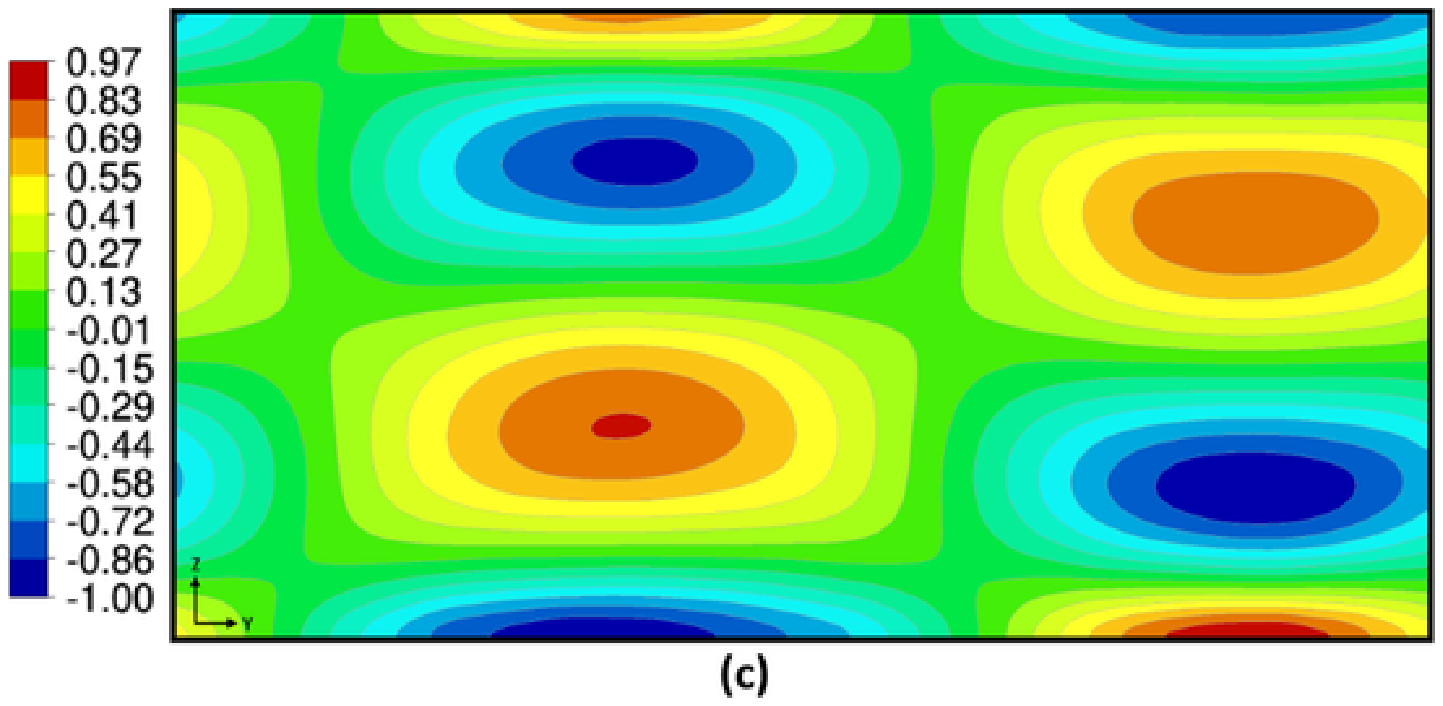}
\caption{\label{Fig14}
The velocity patterns at the stationary stage (a),
the potential temperature deviations $\tilde \Theta \, {\rm Ra}_{_{T}}$ from the equilibrium potential
temperature in the basic reference state (b) and
the total vertical gradient of the mean potential temperature $(\nabla_z \tilde \Theta -1) \, {\rm Ra}_{_{T}}$ (c)
in $yz$ plane for the no-slip vertical boundary conditions for $\epsilon=2.5 \times 10^{-3}$
and the effective Rayleigh number ${\rm Ra}_{_{T}}=1800$.
All fields are normalized by their maximum values.
}
\end{figure}

In this section we use the no-slip boundary conditions for the velocity field in the vertical direction:
\begin{eqnarray}
\tilde {\bm U}(t, z=0) = \tilde {\bm U}(t, z=1)=0 .
\label{B43}
\end{eqnarray}
For the classical laminar convection with the no-slip boundary conditions
[where there are no terms $\propto \epsilon$ in Eq.~(\ref{B2})],
the critical Rayleigh number required for the excitation of convection
is ${\rm Ra}_{\rm cr} \approx 1708$ \cite{EGKR06,CH61,DR02,RH58}.

Let us discuss the results of the mean-field numerical simulations for the case
which takes into account the modification of the turbulent heat flux by the nonuniform motions
[described by the terms $\propto \epsilon$ in Eq.~(\ref{B2})].
In Figs.~\ref{Fig10}--\ref{Fig12} we plot time evolution of the maximum velocity $\tilde {U}_{\rm max}(t)$
for different effective Rayleigh numbers varying from  ${\rm Ra}_{_{T}}=0.5$
to $1800$ and different values of the parameter $\epsilon$.
The large-scale convective-wind instability is excited when the parameter $\epsilon \leq 1.75 \times 10^{-3}$.
After the exponential growth of the maximum velocity $\tilde {U}_{\rm max}(t)$, there is
a saturation stage of the convective-wind instability.
We also observe the nonlinear oscillations of the maximum velocity $\tilde {U}_{\rm max}(t)$
which follow after the stationary stage when the maximum velocity $\tilde {U}_{\rm max}(t)$ reaches the constant.
The characteristic duration of the stationary stage decreases with increase of
the effective Rayleigh number ${\rm Ra}_{_{T}}$ (see Fig.~\ref{Fig10}).
As well as for the stress-free vertical boundary conditions,
the maximum velocity $\tilde {U}_{\rm max}(t)$ in saturation depends strongly on the scale separation parameter
$\epsilon$ (see Figs.~\ref{Fig11} and~\ref{Fig12}), and it is nearly independent of  the effective
Rayleigh number (see Fig.~\ref{Fig10}).

We also show the velocity patterns (Figs.~\ref{Fig13}a and~\ref{Fig14}a) in $yz$ plane at the stationary stage,
the potential temperature deviations $\tilde \Theta \, {\rm Ra}_{_{T}}$ from the equilibrium potential
temperature in the basic reference state (Figs.~\ref{Fig13}b and~\ref{Fig14}b) and
the total vertical gradient of the mean potential temperature $(\nabla_z \tilde \Theta -1) \, {\rm Ra}_{_{T}}$
(Figs.~\ref{Fig13}c and~\ref{Fig14}b).
These figures are for the no-slip vertical boundary conditions at the effective Rayleigh number ${\rm Ra}_{_{T}}=1800$,
and for two values of the parameter $\epsilon=1.75 \times 10^{-3}$ (Fig.~\ref{Fig13})
and $\epsilon=2.5 \times 10^{-3}$ (Fig.~\ref{Fig14}).

As well as for the stress-free vertical boundary conditions, the large-scale circulations form
even below the threshold of the laminar convection for the case of the no-slip boundary conditions.
However, for the no-slip vertical boundary conditions we see some differences.
In particular, increasing the parameter $\epsilon$ from $\epsilon=1.75 \times 10^{-3}$ to
$\epsilon=2.5 \times 10^{-3}$, we observe a transition from the four-cell flow patterns (see Fig.~\ref{Fig13}a,b)
to the two-cell patterns (see Fig.~\ref{Fig14}a,b).
As well as for the stress-free vertical boundary conditions,
Figs.~\ref{Fig13}c and~\ref{Fig14}c also show the regions with the positive total gradient
of the potential temperature $(\nabla_z \tilde \Theta -1) \, {\rm Ra}_{_{T}}$,  which correspond to the stably stratified flows.
This study demonstrates that dependence on the vertical boundary conditions (the stress-free or no-slip boundary conditions)
is not essential.

\section{Discussion}
\label{sec-V}

Let us discuss the novelty aspects and significance of the obtained results.
The effect of modification of the
turbulent heat flux by nonuniform large-scale motions
in turbulent convection was investigated
analytically in Ref.~\cite{EKRZ02},
where the turbulent heat flux ${\bf F}$
given by Eq.~(\ref{R100}) was derived.
This new effect causes an excitation of the convective-wind instability.
The estimate for the growth rate of the convective-wind instability
[see Eq.~(\ref{R200})] was obtained for the case
when the effective Rayleigh number vanishes,
${\rm Ra}_{_{T}} \to 0$ \cite{EKRZ02}.
The linear stage of the convective-wind instability
was numerically studied in Ref.~\cite{EGKR06}
for different conditions.
Applications of these results to atmospheric turbulence
were discussed in Ref.~\cite{EKRZ06},
where the theoretical predictions \cite{EKRZ02}
were compared with observational
characteristics of the cloud convective cells.

In the present paper, we have generalized Eq.~(\ref{R200}) for the growth rate
of the convective-wind instability for
arbitrary effective Rayleigh numbers [see Eq.~(\ref{C1})].
This allows us to obtain equation for the critical effective Rayleigh number
required for the excitation of the large-scale convective-wind instability [see Eq.~(\ref{RC1})].
This equation explains why
the large-scale circulation can be formed in turbulent convection even for very
low effective Rayleigh numbers.
This result has been confirmed by the mean-field simulations
performed in the present study.

We have shown that the convective-wind instability strongly depends
on the scale separation parameter $\epsilon$ characterizing the separation of scales
between the height of the convective layer $L_z$ and the integral turbulence scale $\ell_0$.
In the mean-field numerical simulations,
the parameter $\epsilon$ varies in the range $(0.5 - 2.5) \times 10^{-3}$,
which  corresponds to variations of the separation of scales $L_z/\ell_0$ in the range 12 - 26.
Any mean-field theory is usually valid when the integral turbulence scale is much less
than the characteristic scale of the mean-field variations, that is consistent with
these values $L_z/\ell_0$.
Also direct measurements in laboratory experiments in turbulent convection
by measuring of the two-point correlation function of velocity fluctuations
(which allow to determine the integral scale of turbulence)
are consistent with these values of  $L_z/\ell_0$ (see Refs. ~\cite{BEKR09,BEKR11}).

Since direct numerical simulations (DNS) cannot be performed for very large Reynolds numbers
(i.e., for Reynolds numbers based on integral scale and maximum turbulent velocity
which are larger than $10^4$), while in many applications, e.g., in
atmospheric and astrophysical turbulent flows characteristic Reynolds numbers are much
larger than $10^4$. In this case mean-field simulations based on
nonlinear mean-field equations [where turbulent effects are described
by means of effective (turbulent) transport coefficients
and modified turbulent heat flux] can be very useful.
The mean-field numerical simulations demonstrate existence of the local regions with the positive
vertical gradient of the potential temperature inside the large-scale circulations.
In the present study we explain this effect
[see Eqs.~(\ref{PK1})--(\ref{DDD1}) and corresponding discussion after these equations].
This effect was previously observed in laboratory experiments (see, e.g., \cite{NSS01,BEKR20})
as well as in DNS of turbulent convection (see, e.g., \cite{AB16,KRB17,KA19}).

In view of applications, the obtained results are relevant to
large-scale convective cells (the cloud cells) observed in the atmospheric turbulent convection without strong mean wind.
They are formed in a convective boundary layer with a depth of about $1$ to $3$ km,
and have aspect ratios $L_{z} / L_{\rm hor} \approx  0.05 - 1$
(see, e.g., Ref.~\cite{AZ96}),
where $L_{\rm hor}$ and $L_{z}$ are the horizontal and vertical sizes of convective cells.
The ratio of the minimum size of the convective cells to the maximum scale
of turbulent motions is $L_{\rm min} / \ell_{0} = 5 - 20$.
This implies that the parameter $\epsilon$ for the observed convective cells ranges from $10^{-3}$ to $10^{-2}$.
The characteristic time of formation of the convective cells ($\tau_{\rm form} \sim \tau_0 / \gamma_{\rm
inst}$)  in the atmospheric turbulent convection varies from $1$ to $3$ hours.

Turbulent velocity $u_0$ at the lower part of the surface convective layer,
where the turbulence production is mainly due to the large-scale shear motions,
is of the order of $u_0 \sim (2 - 4) \, u_\ast$, where $u_\ast$ is the friction velocity.
At the upper part of the surface convective layer,
where production of the turbulence is mainly due to the buoyancy,
is of the order of the turbulent convective velocity $u_c = (\beta F_z^\ast \ell_{0})^{1/3}$
(see, e.g., Ref.~\cite{MY75}), which implies that $u_0 /u_c \sim 1$.

\section{Conclusions}
\label{sec-VI}

In the present paper, we study
formation and nonlinear evolution of large-scale circulations
in turbulent convection  by means of the mean-field numerical simulations.
We use periodic horizontal boundary conditions,
and stress-free or no-slip vertical boundary conditions.
We have taken into account the effect of strong modification
of the turbulent heat flux by nonuniform large-scale motions
which generate anisotropic velocity fluctuations.
The performed mean-field numerical simulations have shown that this effect
strongly reduces the critical Rayleigh number
(based on the eddy viscosity and turbulent temperature diffusivity)
required for onset of the large-scale convective-wind instability and formation of large-scale
semi-organised coherent structures (large-scale circulations).
The onset of this instability and the level of the mean velocity at saturation strongly depend
on the scale separation ratio between the height of the convective layer
and the integral scale of turbulence.
The simulations demonstrate existence of the local regions with the positive
vertical gradient of the potential temperature
inside the large-scale circulations.
The latter implies that these regions are stably stratified.

\medskip
\noindent
{\bf ACKNOWLEDGMENTS}

The authors benefited from stimulating discussions
of various aspects of turbulent convection with A. Brandenburg, F.~H.~Busse,
P.~J.~K\"{a}pyl\"{a} and S.~Zilitinkevich.
This research was supported in part by the Israel Ministry of Science
and Technology (grant No. 3-16516) and the PAZY Foundation of the Israel Atomic
Energy Commission (grant No. 122-2020).

\medskip
\appendix

\section{Derivation of  Eq.~(\ref{B2})}
\label{Appendix:A}

In this Appendix, we derive  Eq.~(\ref{B2}) for the mean potential temperature.
To this end, we use Eq.~(\ref{R100}) for the turbulent heat flux ${\bm F}$ with the additional terms
caused by the non-uniform mean flows.
We determine div ${\bm F}$ by assuming that the non-dimensional total vertical heat flux
$F_z^{\rm tot}=F_z^\ast + \meanU_z \, \meanTheta$ is constant.
This yields
\begin{eqnarray}
&& {\bm \nabla} \cdot {\bm F} = - \kappa_{_{T}} \triangle \meanTheta - \tau_0 \,  \biggl[ \left(F_z^{\rm tot} - \meanU_z \, \meanTheta \right) \, \left({\Delta \over 2}-\nabla_z^2\right) \meanU_z
\nonumber\\
&&
+ {1 \over 2} \, \left(\nabla_z \meanU_x - \nabla_x \meanU_z\right) \, \nabla_x \left(\meanU_z \, \meanTheta \right)
+ {1 \over 2} \,\left(\nabla_z \meanU_y - \nabla_y \meanU_z\right)
  \nonumber\\
&&
\times \nabla_y \left(\meanU_z \, \meanTheta \right)
+ \left(\nabla_z \meanU_z\right) \, \nabla_z \left(\meanU_z \, \meanTheta \right)
\biggr] .
 \label{APB2}
\end{eqnarray}
Using Eqs.~(\ref{LB2}) and~(\ref{APB2}), written in non-dimensional form (see definitions of the non-dimensional variables
and the key parameters in Section~\ref{sec-III}), we obtain Eq.~(\ref{B2}).

\section{Derivation of  Eq.~(\ref{C1})}
\label{Appendix:B}

In this Appendix, we derive Eq.~(\ref{C1})
for growth rate of the large-scale convective-wind instability.
To this end, we use linearised non-dimensional equations~(\ref{B1}) and ~(\ref{B2}),
calculate $[{\bm \nabla} \times ({\bm \nabla} \times \tilde{\bm U})]_z$
using the linearised  Eq.~(\ref{B1}) to exclude the pressure term and
seek for solution of the obtained equations in the following form:
$\tilde {\bm U}(t,{\bm x}) = \tilde {\bm U}_0 \exp [{\rm i} \, (\gamma_{\rm inst} t + {\bm K} \cdot {\bm x})]$
and $\tilde \Theta(t,{\bm x}) = \Theta_0 \exp [{\rm i} \, (\gamma_{\rm inst} t + {\bm K} \cdot {\bm x})]$.
This yields the following system of equations:
\begin{eqnarray}
&& \left(\gamma_{\rm inst} + K^2\right)  \,\tilde U_z + {\rm Ra}_{_{T}} \, \left({K_z^2 \over K^2} -1\right) \, \tilde \Theta =0 ,
\label{APB3}\\
&& \left[1 + {\sigma  \over \epsilon \, {\rm Ra}_{_{T}}} \left(K_z^2 - {K^2 \over 2}\right) \right] \,\tilde U_z
- \left(\gamma_{\rm inst} + K^2\right) \,\tilde \Theta =0 ,
\nonumber\\
\label{APB4}
\end{eqnarray}
where $\sigma = 3 \, (u_c/ u_{0})^3$, and we consider, for simplicity, the case when the turbulent Prandtl number
${\rm Pr}_{_{T}}=1$.
Equations~(\ref{APB3}) and~(\ref{APB4}) yield the non-dimensional growth rate~(\ref{C1}) of the large-scale convective-wind instability,
where $K_z = \pi$ and $K\equiv  |{\bm K}| =\pi \, (1 + \alpha^2)^{1/2}$.

\bigskip
\noindent
{\bf DATA AVAILABILITY}
\medskip

The data that support the findings of this study are available from the corresponding author
upon reasonable request.

\bigskip
\noindent
{\bf AUTHOR DECLARATIONS}

\noindent
{\bf Conflict of Interest}

The authors have no conflicts to disclose.

\end{document}